\documentclass[prl,superscriptaddress,twocolumn,floatfix]{revtex4-1}

\usepackage[FIGTOPCAP]{subfigure}
\usepackage[usenames]{color}
\usepackage{amsmath,amssymb}
\usepackage[pdftex]{hyperref,graphicx}
\hypersetup{colorlinks = true, urlcolor = blue, linkcolor = blue, citecolor = blue}

\begin{document}
\title{Inversion-protected higher order topological superconductivity in monolayer WTe$_2$}
\author{Yi-Ting~Hsu}
\email{yhsu2@nd.edu}
\affiliation{Condensed Matter Theory Center and Joint Quantum Institute, University of Maryland, College Park, MD 20742, USA}
\author{William~S.~Cole}
\affiliation{Condensed Matter Theory Center and Joint Quantum Institute, University of Maryland, College Park, MD 20742, USA}
\author{Rui-Xing~Zhang}
\affiliation{Condensed Matter Theory Center and Joint Quantum Institute, University of Maryland, College Park, MD 20742, USA}
\author{Jay~D.~Sau}
\affiliation{Condensed Matter Theory Center and Joint Quantum Institute, University of Maryland, College Park, MD 20742, USA}
\date{\today}

\begin{abstract}
Monolayer $\rm WTe_2$, a centrosymmetric transition metal dichacogenide, has recently been established as a quantum spin Hall insulator and found superconducting upon gating. 
Here we study the pairing symmetry and topological nature of superconducting $\rm WTe_2$ with a microscopic model at mean-field level. Surprisingly, we find that the spin-triplet phases in our phase diagram all host Majorana modes localized on two opposite corners. 
Even when the conventional pairing is favored, we find that an intermediate in-plane magnetic field exceeding the Pauli limit stabilizes an unconventional equal-spin pairing aligning with the field, which also hosts Majorana corner modes. 
Motivated by our findings, we obtain a recipe for two-dimensional superconductors featuring "higher-order topology" from the boundary perspective: 
Generally a superconducting inversion-symmetric quantum spin Hall material whose normal-state Fermi surface is away from high-symmetry points, such as gated monolayer $\rm WTe_2$, hosts Majorana corner modes if the superconductivity is parity-odd.
We further point out that this higher-order phase is an inversion-protected topological crystalline superconductor and study the bulk-boundary correspondence. 
Finally, we discuss possible experiments for probing the Majorana corner modes. 
Our findings suggest superconducting monolayer $\rm WTe_2$ is a playground for higher-order topological superconductivity, and possibly the first material realization for inversion-protected Majorana corner modes \emph{without} utilizing proximity effect. 
\end{abstract}

\maketitle

\textit{Introduction---}
Extensive experimental and theoretical effort has been devoted to transition metal dichalcogenides (TMD), a family of materials with chemical formula $\rm{MX_2}$ (M $=$ transition metal, X $=$ S, Se, Te) known to host a rich variety of intriguing ground states, such as topological insulators and semimetals\cite{qian_2014sci,WTe2QSH_ARPES,Exp_WTe2QSH_Cobden,Exp_WTe2QSH_Pablo,Exp_WTe2WeylNcomm,Exp_WeylMoTe2_nphys}, charge density waves\cite{Thy_CDWTMD_Neto,Exp_TMDCDW_PRL,Exp_NbSe2CDW,Exp_TaS2CDW_Nmat,Exp_CDWNbSe2_Mak,Exp_TaS2CDW_nphys,Exp_GateTunedTiSe2_Neto}, and various types of possibly unconventional superconductivity\cite{Toposc_MoS2_Law,Toposc_holeMoS2_Hsu,Exp_MoS2sc_Iwasa,Exp_IsingSc_Ye,Exp_TMDsc_Iwasa,Exp_NbSe2sc_Mak,Exp_GateTunedTiSe2_Neto,Exp_WTe2sc_Cobden,
Exp_WTe2sc_Pablo}.
Moreover, tuning among these phases is possible by widely accessible experimental knobs, for example changing the thickness, pressure\cite{Exp_TiSe2sc_pressure,Exp_TaS2CDW_Nmat,Exp_WTe2_pressure,Exp_WTe2Pressure_Pan,Exp_MoS2sc_Pressure}, 
electrostatic gating\cite{Exp_MoS2sc_Iwasa,Exp_GateTunedTiSe2_Neto,Exp_TaS2gate}, and recently even the twist angle between monolayers\cite{WSe2_Moire_Wu,MoireHomo_MoTe2_Wu}.
Recently, a centrosymmetric member of the TMD family, monolayer $\rm WTe_2$, has been 
established\cite{qian_2014sci,Exp_WTe2QSH_Cobden, WTe2QSH_ARPES, jia_2017prb, peng_2017ncomm, Exp_WTe2QSH_Pablo} as a quantum spin Hall (QSH) insulator\cite{QSH_KaneMele,QSH_Bernevig}. 
Remarkably, in this same material, superconductivity at temperatures around 1K was soon after reported under tunable electrostatic gating\cite{Exp_WTe2sc_Cobden,Exp_WTe2sc_Pablo}. 
We are thus motivated to understand the nature of this superconductivity
given the prevailing expectation that inducing superconductivity in already topological materials is a promising route for achieving topological superconductors.

Theoretically a known necessary condition for two-dimensional (2D) time-reversal topological superconductors requires negative pairing potentials on an odd number of Fermi surfaces that enclose time-reversal invariant momenta (TRIMs)\cite{IndexTRsc_Zhang,TscMajo_FanZhang}. 
The presence of the inversion symmetry, however, enforces two-fold degeneracy of the Fermi surfaces 
and thus sets up a ``no-go" theorem that precludes such 
superconductors from being topological.  
Nonetheless, recent developments suggest that inversion can unexpectedly enrich the topological structure of a system\cite{Indicator_PRX,InvHOTsc_Khalaf,Z4indicator_Tsc}, and enable new topological crystalline superconductors (TCsc) that are completely beyond the previous paradigm\cite{IndexTRsc_Zhang,TscMajo_FanZhang}. 
In particular, there exists a type of inversion-protected TCsc in dimension $d$ that has no Majorana boundary modes in $d-1$ dimension, yet is still topologically distinct from a trivial superconductor\cite{InvHOTsc_Khalaf,Z4indicator_Tsc}. 
This suggests the possibility that such inversion-protected TCsc belongs to the so-called ``higher order topological phases''\cite{MultipoleTI_science,ZhangMCS, WangMCS,weakHOTsc_Hughes,HOTI_Neupert,MirrorHOTI_Brouwer,
TCsc_Ryu,weakHOTsc_Hughes,HybridHOTsc_Zaletel,HOTIaxion,HingeIronsc_PRL,HOTsc_hetero}, and may host Majorana boundary modes in $d-2$ or lower dimension. 

Here, we propose a surprisingly simple recipe for this exotic inversion-protected TCsc: (1) the normal state is an inversion-symmetric QSH material with Fermi pockets away from TRIMs, and (2) the superconductivity is parity-odd. 
Given that gated monolayer $\rm WTe_2$ readily satisfies criteria (1), unconventional superconductivity with odd parity becomes the last piece of the puzzle for an inversion-protected TCsc that could host exotic Majorana corner modes. 

In fact, in $\rm WTe_2$ there is ample reason to suspect that electron correlations might be strong, and odd-parity superconductivity is therefore plausible. First is the fact that the reported superconductivity\cite{Exp_WTe2sc_Pablo, Exp_WTe2sc_Cobden} 
occurs at a low carrier density, while \textit{ab initio} calculations do not reproduce the low-energy normal state band structure found by angle-resolved photoemission spectroscopy (ARPES)\cite{WTe2QSH_ARPES} and scanning tunnelling microscopy (STM)\cite{WTe2QSH_ARPES} studies unless one goes beyond the generalized-gradient approximation 
\cite{qian_2014sci,DFTHybridFunc,WTe2QSH_ARPES}. Moreover, the reported in-plane upper critical field $H_{c2}^{\parallel}$ is 2.5-4.5 times higher than the Pauli limit $H_p$\cite{Exp_WTe2sc_Pablo, Exp_WTe2sc_Cobden}. While an $H_{c2}^{\parallel}$ higher than the Bardeen-Cooper-Schrieffer theory prediction in centrosymmetric materials can occur when the normal state has a high spin-orbit scattering rate\cite{Hcs_SOCrate} or when the g factor deviates from two\cite{Exp_WTe2sc_Cobden}, another possible origin is a spin-triplet (and thus odd-parity) paired state with spin aligning in the field direction.

In this work, we report the pairing symmetry and topological nature of the newly discovered  superconductivity in gated monolayer WTe$_2$. 
First, we solve the linearized gap equations to obtain a superconducting phase diagram in terms of microscopic interactions. 
By investigating the boundary modes in different phases,  
we find Majorana corner modes in odd-parity phases and surprisingly, also in the field-induced equal-spin phase emerging upon the suppression of conventional pairings.
Then, we obtain a general recipe from the boundary perspective for achieving such 2D superconductivity with corner Majoranas.
Finally, we point out that such higher-order phase is an inversion-protected topological crystalline superconductor that can be characterized by a bulk invariant we propose, and address the bulk-boundary correspondence.   
Our recipe provides a new route towards materializing a novel topological phase of matter, as well as realizing Majorana zero modes, which is the first step for topological quantum computation. 

\begin{figure}[!]
\includegraphics[width=8cm]{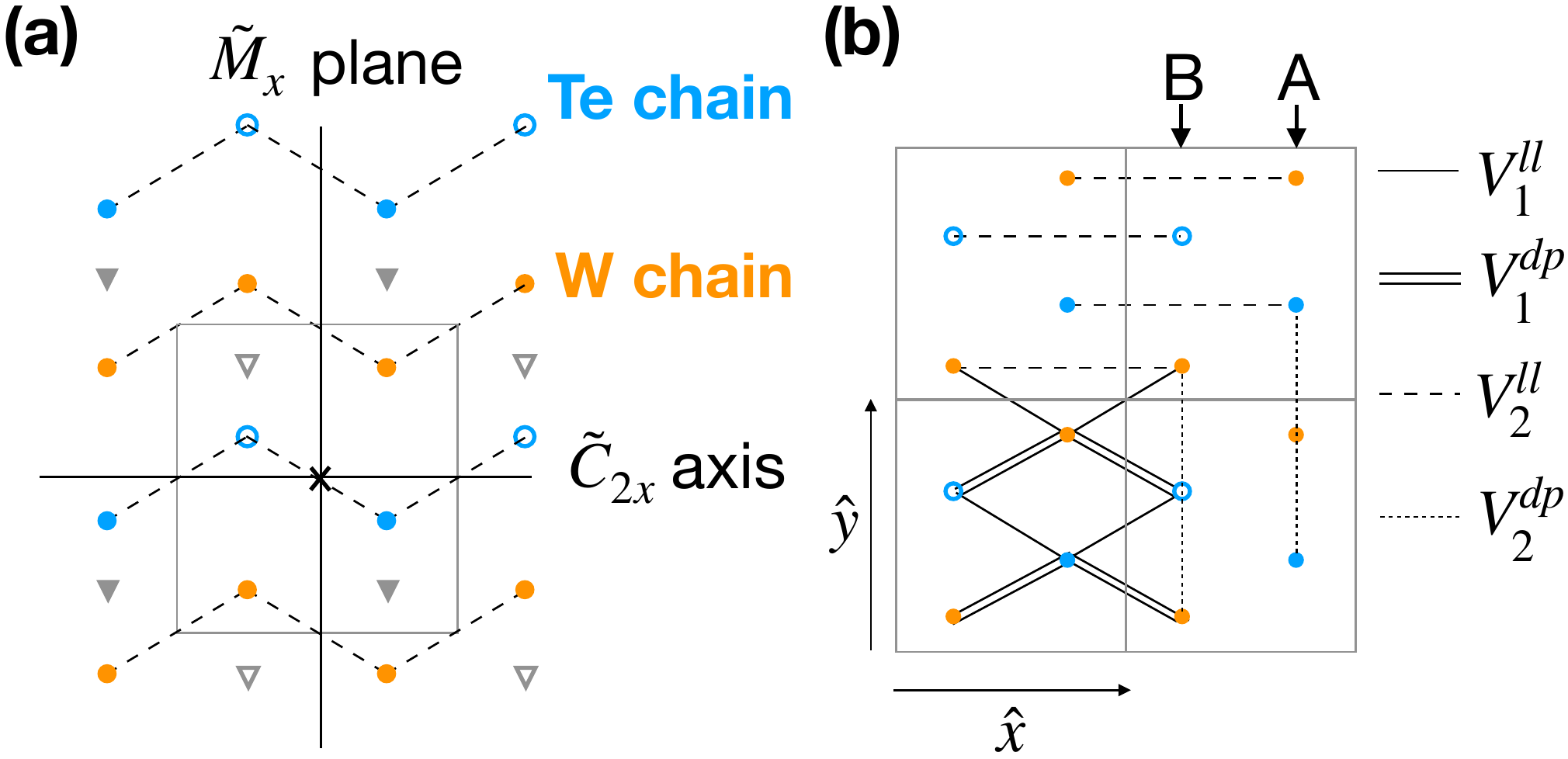}
\caption{
Schematics for (a) the top view of the lattice of $1T^{\prime}$-WTe$_2$, and (b) the microscopic interactions considered in Eq.~\ref{Hint}. In (a), the filled orange circles represent the $\rm W$ atoms, which locate on the $z=0$ plane. The filled and hollow blue circles (grey triangles) represent the $\rm Te$ atoms above and below the $z=0$ plane, which are (are not) associated with the Wannier orbital centers in the low-energy tight-binding description. The grey rectangle indicates a unit cell, 
 the horizontal and vertical black lines show the screw-rotation axis and the glide-mirror plane respectively, and the black cross marks the inversion center. In (b), we omit the $\rm Te$ atoms (grey triangles) that do not contribute to Wannier orbitals.
}
\label{lattice} 
\end{figure}

\textit{Model---} Monolayer WTe$_2$ is stable in the $1T^{\prime}$ structure, which is a buckled honeycomb lattice that is distorted into a rectangular lattice consisting of in-plane and buckled zigzag chains of W and Te atoms, respectively, see Fig. \ref{lattice}(a). 
This lattice is nonsymmorphic, with a two-fold screw rotation $C_{2x}$ 
and a glide mirror symmetry $M_x$\footnote{$M_x$ can be transformed back to the conventional mirror symmetry by shifting the mirror plane away from inversion center.},  
each with a half-unit-cell translation along the chain direction $\hat{x}$. The lattice also has inversion symmetry $I_0$, resulting from the product of the two symmetries. 

To study the dominant pairing channels in gated monolayer WTe$_2$, we start from a minimal tight-binding model previously obtained by other authors from a low-energy fit to \emph{ab initio} calculations\cite{muechler_2016prx, WTe2_tbSOC}\footnote{Ref. \onlinecite{tbmodelmoreSOC} showed that more spin-orbit coupling terms are required for a better fit to the experimentally data. We nonetheless expect our results to change only quantitatively.}. The Hamiltonian is written in a basis of spin s and four Wannier orbitals. These Wannier orbitals are labeled by the sublattice $\sigma=A, B$ they are on and by whether they transform as $d_{x^2-y^2}$ or $p_x$ orbitals ($l=d,p$). The $l=d,p$ orbitals are derived from W and Te atoms, respectively. Each degree of freedom is denoted by the corresponding Pauli matrices, $\hat{s}$, $\hat{\sigma}$, and $\hat{l}$, respectively. 
The full normal-state Hamiltonian is
\begin{align}
H_0(\textbf{k}) = \hat{s}_0 \otimes \left( \hat{h}_0(\textbf{k}) - \mu \right) + V_{soc} \hat{s}_z \hat{\sigma}_z \hat{l}_y. 
\label{H0}
\end{align}
Here, the $s_z$-preserving intrinsic spin-orbit coupling $V_{soc}$\cite{WTe2_tbSOC} is the lowest order term in $\textbf{k}$ that obeys time-reversal, screw rotation, and glide mirror symmetries, while the spin-degenerate part $\hat{h}_0(\textbf{k})$\cite{muechler_2016prx, WTe2_tbSOC} is a $4 \times 4$ matrix in the basis of $\hat{\sigma} \otimes \hat{l}$ {\color{magenta}[Supplementary Material (SM) Sec. I]}.
As a zeroth-order approximation to the gating effects, we set the overall chemical potential $\mu=0.5$. The resulting two electron pockets are centered along the $\Gamma-X$ line [Fig. \ref{PD}(a)], as observed by ARPES\cite{WTe2QSH_ARPES}.

We consider short-ranged density-density interactions that preserve the lattice symmetries up to nearest-neighbor unit cells [Fig. \ref{lattice}(b)]:
\begin{align}\label{Hint}
H_{\rm int} = \sum_{\textbf{r}\textbf{r}^{\prime}} &\sum_{\alpha\beta\alpha^{\prime}\beta^{\prime}}
\Gamma_{\alpha^{\prime}\beta^{\prime},\beta\alpha}(\textbf{r},\textbf{r}^{\prime})
c_{\textbf{r}\alpha^{\prime}}^{\dagger} c_{\textbf{r}^{\prime}\beta^{\prime}}^{\dagger}
c_{\textbf{r}^{\prime}\beta} c_{\textbf{r}\alpha} \nonumber\\
= \sum_{\textbf{r}}
&U^{l} n_{\uparrow\sigma l}(\textbf{r}) n_{\downarrow \sigma l}(\textbf{r})+V^{ll'}_a n_{\sigma l}(\textbf{r}) n_{\sigma' l'}(\textbf{r}+\boldsymbol{\delta}_a)
\end{align}
where $\sigma^{(')}$, $l^{(')}$, and $a=1,2$ indices are summed over, $n_{s\sigma l}(\textbf{r})$ is the density with spin $s$ and orbital $l$ locating at sublattice $\sigma$ in the unit cell centered at $\textbf{r}$, and $n_{\sigma l}(\textbf{r})= \sum_s n_{s\sigma l}(\textbf{r})$. 
Here $U^{l}$ denotes the on-site interactions for orbital $l$, $V_1^{ll'}$ and $V_2^{ll'}$ denote the nearest- and next nearest-neighbor interactions respectively on the zigzag chains with intra (inter)-orbital characters for $l'=l(\bar{l})$ [Fig. \ref{lattice}(b)], and $\boldsymbol{\delta}_a$ denotes corresponding lattice vectors {\color{magenta}[SM Sec. I]}. For simplicity, in the following we consider the case where $U^{l}=U$, and $V_1^{ll'}=V_2^{ll'}=V$.

\begin{table}[t!]
\centering
\begin{tabular}{ |c|c|c|c| }
\hline
  & $\eta_{C_{2x}}$ & $\eta_{M_{x}}$ & Examples \\
 \hline
 $A_g$ &+ &+ & $~~\hat{s}_0\otimes\hat{\sigma}_0\otimes\hat{l}_{0}~~$\\
 \hline
 $B_g$ &- &- & $~~\hat{s}_0\otimes\hat{\sigma}_z\otimes\hat{l}_{x}~~$\\
 \hline
 $A_u$ &+ &- & $~~k_x \hat{s}_x\otimes\hat{\sigma}_0\otimes\hat{l}_{z}~~$\\
 \hline
 $B_u$ &- &+ & $~~k_x \hat{s}_z\otimes\hat{\sigma}_0\otimes\hat{l}_{z}~~$\\
 \hline
\end{tabular}\\
\caption{
The parities of the irreducible representations under the $1T^{\prime}$ lattice symmetry operations. The action of the symmetries on crystal momentum and internal indices and the used Nambu basis are shown in the text.
}
\label{symmetry}
\end{table}

\textit{Method and phase diagram---}
To analyze the dominant pairing channel for given interactions $U$ and $V$, we first classify the symmetries of possible pairing gaps.  
The normal state preserves two nonsymmorphic symmetries $C_{2x}=e^{ik_xa_x/2}(-i\hat{s}_x\otimes \hat{\sigma}_x\otimes \hat{l}_0$), $k_y\rightarrow -k_y$, and $M_x=e^{ik_xa_x/2}(-i\hat{s}_x\otimes \hat{\sigma}_0\otimes \hat{l}_z$), $k_x\rightarrow -k_x$. The mean-field Bogoliubov-de Gennes (BdG) Hamiltonian
\begin{align}
H^{\rm BdG}_{\textbf{k}} = \left(\begin{array}{cc}
H_{0}(\textbf{k}) & \Delta(\textbf{k}) \\
\Delta^\dagger(\textbf{k}) & -T^{\dagger}H_{0}^{\dagger}(\textbf{k})T
\end{array}\right) 
\label{Hbdg}
\end{align}
therefore obeys $g^{BdG}_{\textbf{k}}H^{\rm BdG}_{\textbf{k}}(g^{BdG}_{\textbf{k}})^{\dagger}=H^{\rm BdG}_{g\textbf{k}}$, where 
$T=is_y\mathcal{K}$, $\textbf{k}\rightarrow -\textbf{k}$ is the time-reversal operation with $\mathcal{K}$ the complex conjugation, and $g^{BdG}_{\textbf{k}}=\rm diag$ $[g_{\textbf{k}},\eta_gg_{\textbf{k}}]$ describes how the two symmetries $g=C_{2x},M_x$  act on the Nambu basis $[c_{\textbf{k}\uparrow},c_{\textbf{k}\downarrow},c^{\dagger}_{\textbf{k}\downarrow},-c^{\dagger}_{\textbf{k}\uparrow}]$.
Thus, the pairing gaps transform as $g_{\textbf{k}}\Delta_{\textbf{k}}g^{\dagger}_{\textbf{k}}=\eta_g\Delta_{\textbf{k}}$, and we can classify all possible pairing gaps into four irreducible representations $A_g$, $B_g$, $A_u$, and $B_u$ according to their parities $\eta_g=\pm 1$ under the symmetry transformations $g$, see Table \ref{symmetry}. 

Next, we determine which irreducible representation has the highest $T_c$ by solving the linearized gap equation\cite{ScLectureSigrist} 
$\Delta_{\alpha^{\prime}\beta^{\prime}}(\textbf{k}^{\prime})=-\sum_{\textbf{k}^{\prime\prime}\textbf{k}}\Gamma_{\alpha^{\prime}\beta^{\prime},\beta''\alpha''}(\textbf{k}^{\prime},\textbf{k}'') \times 
\chi_{\beta''\alpha'',\alpha\beta}(\textbf{k}'',\textbf{k},T)\Delta_{\alpha\beta}(\textbf{k})$, 
where Greek indices contain all the internal indices $(s,\sigma,l)$, and repeated indices are summed over. Here, the interaction 
$\Gamma_{\alpha^{\prime}\beta^{\prime},\beta\alpha}(\textbf{k}^{\prime},\textbf{k})$ is the Fourier transform of $\Gamma_{\alpha^{\prime}\beta^{\prime},\beta\alpha}(\textbf{r},\textbf{r}^{\prime})$ in Eq. \ref{Hint}, and $\chi_{\beta''\alpha'',\alpha\beta}(\textbf{k}'',\textbf{k},T)$ is the non-interacting static pairing susceptibility at temperature $T$. Solving 
the lineared gap equation amounts to solving the eigenvalue problem of the effective interaction projected onto the Fermi surface $\tilde{\Gamma}(\textbf{p}^{\prime},\textbf{p})=-\sqrt{P}_{\textbf{p}^{\prime}}\Gamma(\textbf{p}^{\prime},\textbf{p})\sqrt{P}_{\textbf{p}}$, 
where $\textbf{p}^{(\prime)}$ is the incoming (outgoing) momentum on the Fermi surface, and $P_{\textbf{p}}=(\sum_{n=1,2}|\textbf{p},n\rangle\langle\textbf{p},n|)\otimes(\sum_{n=1,2}|-\textbf{p},n\rangle\langle-\textbf{p},n|)$ projects an electron-pair state to the two degenerate non-interacting bands $n$ on the Fermi surface at momenta $\textbf{p}$ and $-\textbf{p}$. The eigenvector $\psi(\textbf{p})$ of $\tilde{V}$ with the most negative eigenvalue $\lambda$ is the solution to the linearized gap equation 
with the highest $T_c\propto \exp(-1/|\lambda|)$. We can then determine how $\psi(\textbf{p})$ behaves under symmetries $C_{2x}$ and $M_x$ under different interactions and obtain the superconducting phase diagram of $H=H_0+H_{\rm int}$. 

\begin{figure}[!]
\includegraphics[width=8cm]{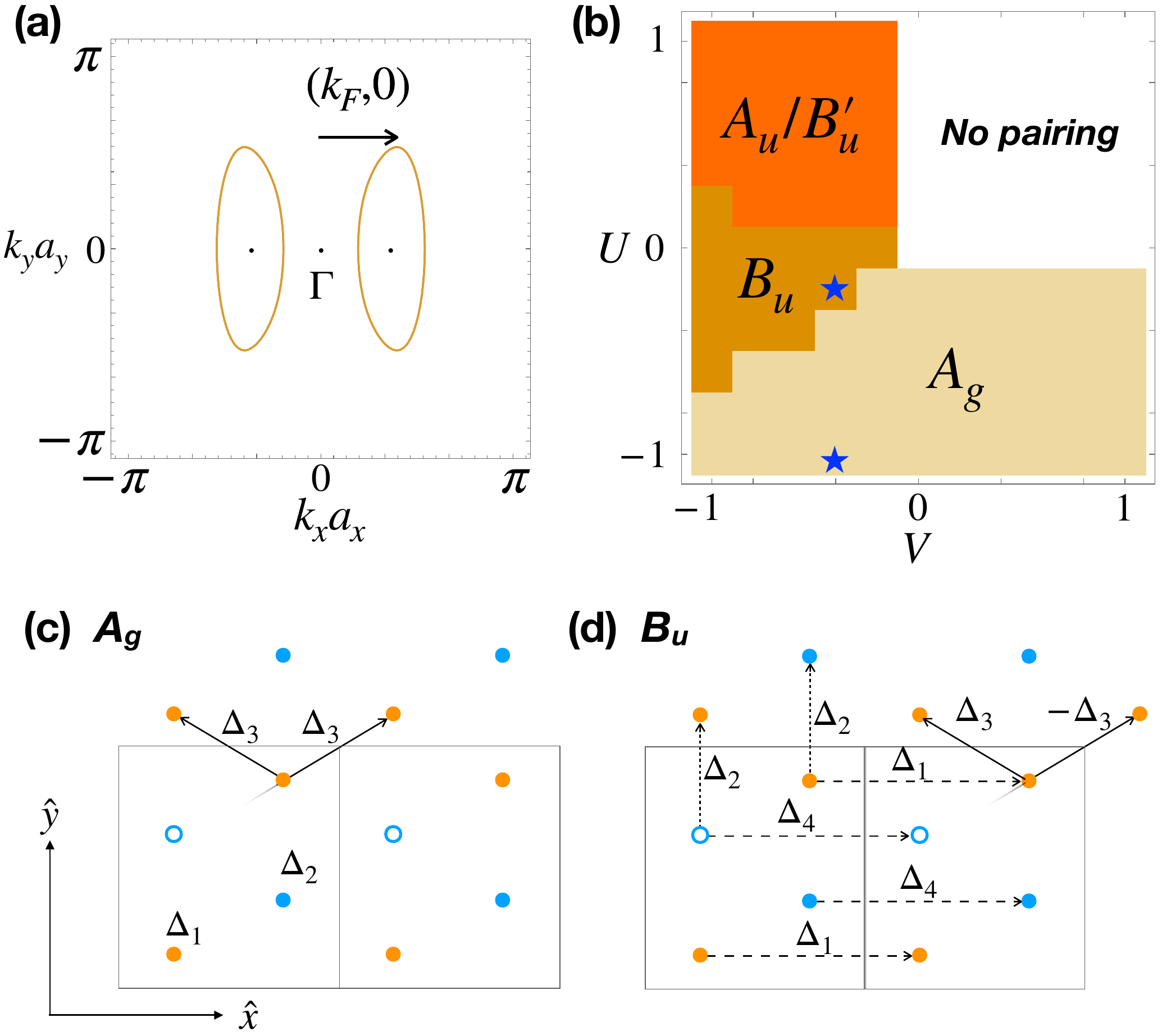}
\caption{
(a) The two Fermi pockets of $H_0$ at chemical potential $\mu=0.5$. $a_x$ and $a_y$ are the lattice constants of a unit cell. 
(b) Phase diagram obtained from solving the linearized gap equation. The blue stars mark the representative points we study for even- and odd-parity pairings in the rest of the paper. 
The spatial configurations of the dominant components in the self-consistent solutions with (c) $A_g$ and (d) $B_u$ symmetries, computed on a system with $12$ by $12$ unit cells. $\Delta_i\equiv|\Delta_{\alpha'\beta'}(\textbf{r},\textbf{r}')|$ for the bond with the  $i^{\rm{th}}$ largest gap magnitude. $\Delta_{1/2}$ in (c) denotes the magnitude for on-site gaps. $A_u$ and $B_u'$ have similar configurations to that of $B_u$ despite different spin structures. 
} 
\label{PD}
\end{figure}

In Fig. \ref{PD}(b) we present this phase diagram as a function of $U$ and $V$.  
We find that while on-site attractions favor the even-parity ``trivial'' representation $A_g$ as expected, the odd-parity representations $A_u$ and $B_u^{(')}$ (the superscript denotes different pair spin textures) dominate over a large portion of the phase diagram where the nearest-neighbor attraction $V$ dominates. 
In particular, the degenerate $A_u$ and $B_u'$ gaps at repulsive $U$ are equal-spin triplet in the out-of-plane direction ($|\uparrow\uparrow\mp\downarrow\downarrow\rangle$), and the $B_u$ gap at attractive $U$ has $s_z=0$ ($|\uparrow\downarrow+\downarrow\uparrow\rangle$). 
This $SU(2)$-symmetry breaking is due to the intrinsic spin-orbit coupling $V_{soc}$. 

We can understand qualitatively the competition between even- and odd-parity pairings from their real-space gap structures. 
To this end, we write down the mean-field Hamiltonian in Eq. \ref{Hbdg}
in real space and solve the self-consistency equations $\Delta_{\alpha'\beta'}(\textbf{r},\textbf{r}')=-\sum_{\alpha\beta}\Gamma_{\alpha'\beta',\beta\alpha}(\textbf{r},\textbf{r}')\langle c_{\textbf{r}'\beta}c_{\textbf{r}\alpha}\rangle$ by iteration.
We consider the short-ranged interactions $H_{\rm int}$, and show results for representative points for even- and odd-parity pairings [see blue stars in Fig.~\ref{PD}(b)].   
We find the dominant component in the even-parity $A_g$ gap to be the on-site pairings as expected, 
while the dominant contribution to the odd-parity $B_u$ gap comes from the next-nearest-neighbor d-orbital pairing along the chains in the $\hat{\textbf{x}}$ direction [see the bonds with $\Delta_1$ in Fig.~\ref{PD}(c) and (d)]. 
It is then clear that attractions $U^d$ and $V_2^{dd}$ in Eq. \ref{Hint} are the main terms responsible for $A_g$ and $B_u$ pairings respectively. 
While on-site attractions $U^d$ are uniform in momentum space and promote even-parity pairing, attractive $V_2^{dd}$ terms enhance scattering processes with large momentum-transfer $2k_F$ across the two pockets, which promotes odd-parity pairing {\color{magenta}[SM Sec. II]}; hence the balance between even- and odd-parity pairings as shown in the phase diagram.  

\textit{Corner Majoranas in \rm{WTe}$_2$---} 
To understand the topological properties of these phases, we examine the boundary modes of different paired states in the phase diagram. 
While the spin-singlet $A_g$ phase is topologically trivial as expected, we find that spin-triplet phases exhibit exotic boundary modes. 
Specifically, in our model for superconducting WTe$_2$ given by $H_0$ and the self-consistently obtained $B_u$ pairing
\footnote{For numerical convenience we take the self-consistent $B_u$ symmetry solution and multiply by 10, so that the resulting superconducting gaps are always much larger than any finite-size gaps of the normal bulk or edge states for tractable lattice sizes.}
, we numerically demonstrate the existence of zero-energy corner-localized states on an open-boundary geometry [Fig.~\ref{fig:mcs}].    
We further verify that with an increasing system size $L$, these corner states tend exponentially toward zero energy {\color{magenta}[SM Sec. III]}, which unambiguously demonstrate the existence of Majorana Kramers pairs localized at two opposite corners. 
We also find similar Majorana corner modes the other spin-triplet phase $A_u$ {\color{magenta}[SM Sec. III]}. 

Even if the realistic WTe$_2$ lies in the even-parity pairing $A_g$ regime in Fig. \ref{PD}(b), we find that an intermediate in-plane magnetic field can surprisingly drive a first-order phase transition and stabilize a new equal-spin phase $B_u''$\footnote{The superscript denotes a different spin orientation from that of $B_u^{(')}$.} aligning with the applied field near the Pauli limit [Fig.~\ref{hx}(a)]. 
This is consistent with the in-plane critical field exceeding the Pauli limit reported by recent experiments\cite{Exp_WTe2sc_Pablo, Exp_WTe2sc_Cobden}. 
Importantly, this field-induced $B_u''$ phase also exhibits two majoranas localized near opposite corners [Fig.~\ref{hx}(b)(c)]. Due to the broken time-reversal symmetry, these two corner modes are single Majoranas instead of Majorana Kramers pairs {\color{magenta}[SM Sec. III]}. 
We therefore emphasize that even if the superconductivity in the realistic WTe$_2$ belongs to the even-parity $A_g$ representation, it is stil possible to obtain single Majorana corner modes by applying an in-plane field. 
\begin{figure}[t]
    \includegraphics[width=8cm]{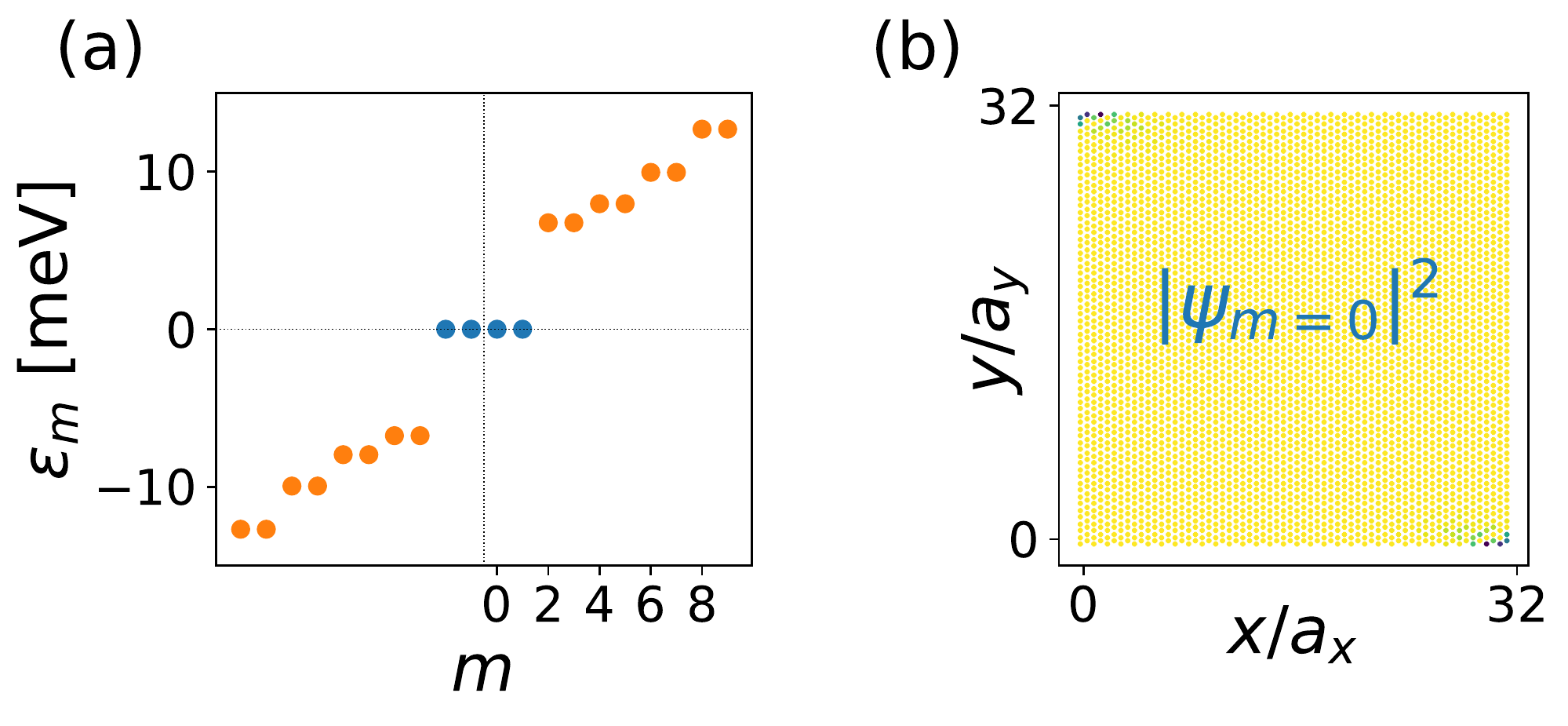}
    \caption{BdG spectrum for gated WTe$_2$ with $B_u$ pairing symmetry at $U=-0.2$, $V=-0.4$ on a finite lattice of $32 \times 32$ unit cells computed by Lanczos techniques.
    (a) The gapped BdG spectrum with a near-zero energy Majorana-Kramers' doublet.
    (b) The spatial probability distribution $|\psi_0|^2$ corresponding to the zero modes, demonstrating sharp corner localization. The geometry preserves inversion, but not the two nonsymmorphic symmetries.}
    \label{fig:mcs}
\end{figure}

\textit{A recipe for 2D higher-order superconductors---}\\
We point out that these 2D higher-order superconducting phases can in fact be achieved by a general recipe\footnote{There could be more recipes to achieve such a phase.}: 
Our studies on WTe$_2$ suggests that corner majoranas might occur generically from the combination of a gated QSH state 
with odd parity superconductivity.
This recipe is most intuitive from the boundary perspective. 
Consider such a QSH normal state at a doping level where it still exhibits counter-propagating modes well-localized on the edge. 
In the absence of pairing, the corresponding BdG Hamiltonian has two electron-like and two hole-like zero-energy eigenstates with edge-localized wavefunctions. 
When we introduce an odd-parity pairing potential, which changes sign in real space when projected onto opposite edges and inevitably vanishes at the domain walls, the electron- and hole-like edge states will mix and acquire finite energies \textit{except} at the two inversion-related points where the projected pairing vanishes. 
The resulting ``leftover'' zero-energy modes, whose point-like wavefunctions will likely to be trapped at corners for realistic samples, therefore lead to two Majorana Kramer's pairs localized on two opposite corners. 
Although the bulk-boundary correspondence is not rigorously proven, we analytically show that corner majoranas naturally exist in a minimal model we construct for superconductors built from our recipe {\color{magenta}[SM Sec. VII]}. 

\begin{figure}[t!]
\includegraphics[width=8cm]{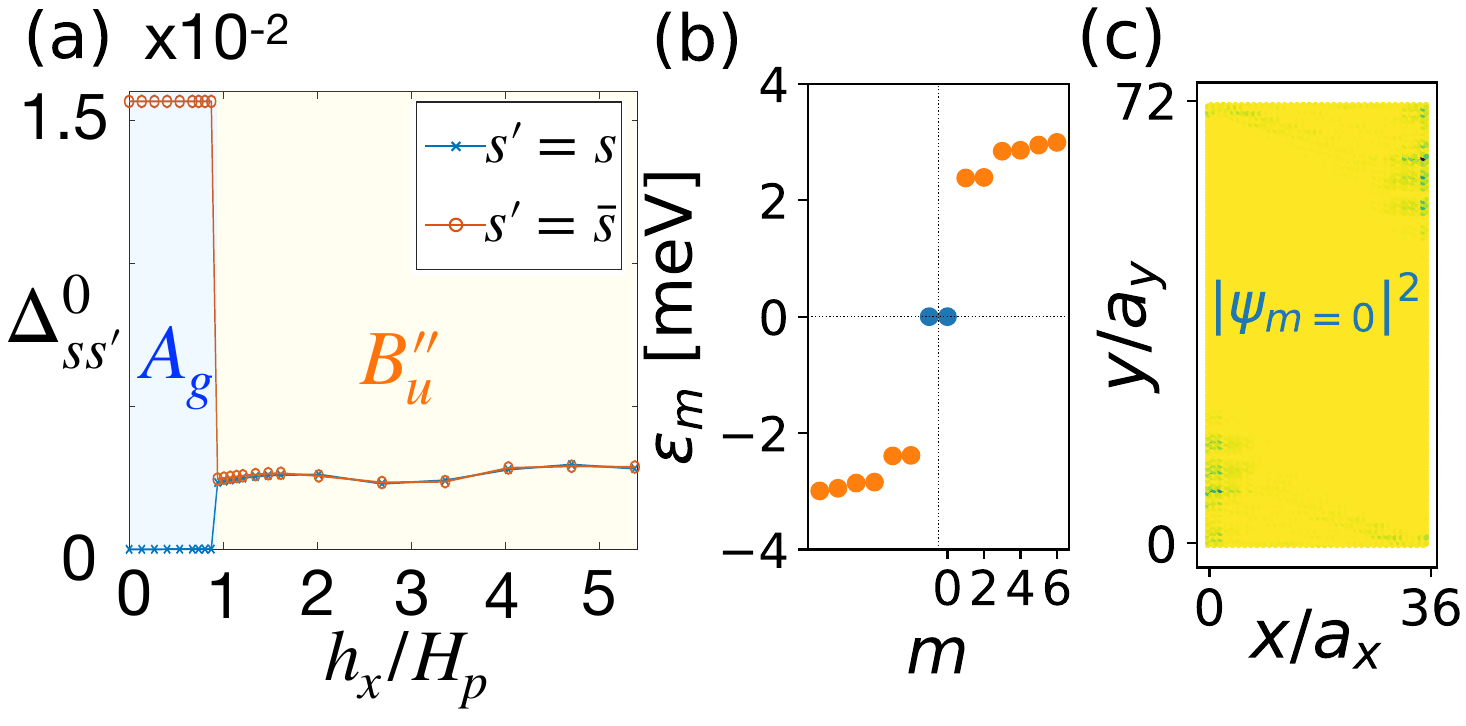}
\caption{
(a) The evolution of pairing symmetries and dominant order-parameter magnitudes of different spin components $\Delta^{0}_{ss}$ and $\Delta^{0}_{s\bar{s}}$ when applying 
an in-plane field with strength $h_x$ to the $A_g$ phase. 
We consider a representative point $(U,V)=(-1,-0.4)$ [the lower blue star in Fig. \ref{PD}(b)], and solve the gap equations self-consistently with term $H_{\rm{field}}=h_x \hat{s}_x \otimes \hat{\sigma}_0 \otimes \hat{l}_0$ added to Eq. \ref{H0}. The blue and yellow background colors represent phase $A_g$ and $B_u''$. For $A_g$, the opposite-spin component results from spin-singlet pairing. For $B_u''$, the opposite- and equal-spin components result from spin-triplet states $|\uparrow\downarrow+\downarrow\uparrow\rangle$ and $|\uparrow\uparrow+\downarrow\downarrow\rangle$ respectively. (b) The gapped BdG spectrum with zero-energy modes, and (c) the probability distribution of the zero-energy eigenstate $|\psi_0|^2$ at $h_x/H_p\sim 5.3$.
}  
\label{hx}
\end{figure}

\textit{Bulk invariant perspective of the recipe---}
This 2D higher-order topological superconducting state is in fact a type of TCsc protected by inversion symmetry. 
Based on studies of various symmetry-protected topological phases\cite{FuKaneZ2,Indicator_PRX,Z4indicator_Tsc}, we conjecture that the bulk topology in inverion-protected TCscs could be inferred from the inversion eigenvalues of occupied BdG bands at TRIMs. 
With these BdG parity data, we define a symmetry indicator as the bulk invariant for a 2D inversion-protected TCsc in the presence of time-reversal symmetry 
\begin{align}
\kappa=\frac{1}{4}\sum_{\textbf{k}\in \rm{TRIM}}\sum_{n}\xi_{\textbf{k}n},  
\label{kappa} 
\end{align}
inspired by indicators proposed for 3D systems\cite{Indicator_PRX,Z4indicator_Tsc}.  
Here $\xi_{\textbf{k},n}$ are the parity eigenvalues of the occupied BdG bands at TRIMs $\textbf{k}$\cite{FuKaneZ2}, and this indicator is stable to adding trivial normal bands for restricted cases where the normal state is half-filled. 
Application of this formula thus requires extending the ``normal'' inversion operator $I_0$ to Nambu space. For odd-parity superconductors, which are defined by superconducting gaps satisfying $I_0 \Delta_\textbf{k} I_0^{-1}=-\Delta_{\textbf{k}}$, the operator $I=\rm diag$$(I_0, -I_0)$ defines the inversion operation for BdG Hamiltonians.  
For even-parity superconductors, this inversion operator $I$ has no minus sign in the hole part, so $\kappa$ is always 0\cite{Z4indicator_Tsc}. 
By identifying trivial BdG parity data as those from ``atomic superconductors'', which are constructed by placing zero-dimensional electron- or hole-like bogoliubons at Wycoff positions, we can see that our indicator is stable upon mod 4. 
This indicates that the classification of 2D inversion-protected TCsc is $Z_4$. 

To identify which of the four states features corner Majoranas, we relate our index $\kappa$ 
to the well-known $Z_2$ index $\nu$ for 2D time-reversal superconductors{\color{magenta}[SM Sec. IV]}: 
\begin{align}
\nu=\kappa ~~\rm mod~~2.      
\label{kappanu} 
\end{align}
It is thus clear that $\kappa=0,2$ phases do not have edge Majoranas while $\kappa=1,3$ phases do. Nonetheless, the $\kappa=2$ phase is topologically distinct from the trivial $\kappa=0$ phase, hinting that the former has corner Majoranas. 

In fact, the 
phases hosting corner Majoranas in WTe$_2$ has $\kappa=2$, which we explicitly verified using $H^{\rm BdG}$ with self-consistently obtained $\Delta$ in Fig. \ref{PD}(d) {\color{magenta}[SM Sec. V]}.
Not only for this particular example, here we show that 
general 2D higher-order superconductors constructed from our recipe 
have $\kappa=2$. 
To see this, we relate the $Z_4$ indicator $\kappa$ for a time-reversal parity-odd BdG system to the $Z_2$ topological index $\nu_N$\cite{KaneMele_Z2} for its normal state:   
\begin{align}
\kappa=2\kappa_N,~~~~\nu_N=\kappa_N.     
\label{kappanu} 
\end{align}
Here, $\kappa_N=0,1$ is the $Z_2$ indicator defined analogously as in Eq. \ref{kappa} but for normal-state Hamiltonians\cite{Z4indicator_Tsc}. Importantly, the latter relation holds for a metallic state only when the numbers of occupied bands are all the same at all TRIMs {\color{magenta}[SM Sec. VI]}. 
Now, we follow our recipe and take the normal state to be a gated QSH state whose FS does not circle any TRIM, just as gated WTe$_2$. In this case, Eq. \ref{kappanu} holds and we have $\kappa_N=\nu_N=1$. Upon introducing odd-parity pairing, the resulting superconductor therefore has $\kappa=2$. 

\textit{Discussion---}
For the odd-parity paired states we find in WTe$_2$, which we find to be inversion-protected higher-order TCsc, 
we expect that the Majorana corner modes cannot be removed without closing the bulk gap if inversion is preserved. 
When the inversion symmetry is broken, while the Majoranas are no longer protected by the 2D bulk topology, they are still protected by the gaps on the 1D edges. In this case the paired state becomes the so-called ``extrinsic'' higher-order topological superconductor\cite{ExtrinsicHOTsc}.    
We thus expect these Majorana corner modes can in principle be probed by STM or transport measurements.

\emph{Acknowledgment}---This work is supported by Microsoft and Laboratory for Physical Sciences. R.-X.Z. is supported by a JQI Postdoctoral Fellowship. JS was supported by the NSF-DMR1555135 (CAREER). The authors acknowledge the University of Maryland super-computing resources (http://hpcc.umd.edu) made available for conducting the research reported in this paper. 
This research was supported in part (through helpful discussions at KITP) by the National Science Foundation under Grant No. NSF PHY-1748958. 
\textit{Note added--} After posting this work we became aware of Ref. \onlinecite{Indicator_Fischer}, which mainly discussed the formulation of symmetry indicators for inversion-protected TCsc in any $d$ dimension. Their $d=2$ case agrees with our conjecture in Eq. \ref{kappa} for the cases we focus on.

%

\maketitle
\begin{center}
{\bf\normalsize{SUPPLEMENTARY MATERIALS}}
\end{center}

\section{I.~~~Normal state and interacting Hamiltonians}
For the normal state, we adapt the first-principles-derived model reported in Ref.~\onlinecite{WTe2_tbSOC} for the normal-state Hamiltonian $H_0$ [see Eq. 1 in the main text] of monolayer WTe$_2$.  
The spin-degenerate part $\hat{h}_0(\textbf{k})$ in $H_0$ is given by 
\begin{align}
h_0(\textbf{k}) = \left(\begin{array}{cccc}
\varepsilon_d(\textbf{k}) & 0 & t^{AB}_d g_{\textbf{k}} & t^{AB}_0 f_{\textbf{k}} \\
0 & \varepsilon_p(\textbf{k}) & -t^{AB}_0 f_{\textbf{k}} & t^{AB}_p g_{\textbf{k}} \\
t^{AB}_d g^\ast_{\textbf{k}} & -t^{AB}_0 f^\ast_{\textbf{k}} & \varepsilon_d(\textbf{k}) & 0 \\
t^{AB}_0 f^\ast_{\textbf{k}} & t^{AB}_p g^\ast_{\textbf{k}} & 0 & \varepsilon_p(\textbf{k})
\end{array}\right)
\end{align}
in the basis of $\hat{\sigma} \otimes \hat{l}$. 
Here the momentum dependence is contained in the functions $\varepsilon_{l}(\textbf{k}) = \mu_l + 2t_l \cos(k_x) + 2t'_l \cos(2k_x)$, $f_{\textbf{k}} = 1 - e^{-ik_x}$, $g_{\textbf{k}} = (1 + e^{-ik_x})e^{ik_y}$. Following Ref.~\onlinecite{WTe2_tbSOC}, we fix the tight-binding parameters (in eV) as:
$\mu_d = 0.4935, \mu_p = -1.3265, t_d = -0.28, t'_d = 0.075, t_p = 0.93, t'_p = 0.075, t_0^{AB} = 1.02, t_d^{AB} = 0.52, t_p^{AB} = 0.40, V_{soc} = 0.115$. 

For the interacting part, we consider short-ranged density-density interactions with the following explicit form: 
\begin{align}\label{Hint}
H_{\rm int} 
= \sum_{\textbf{r}}
&U^{l} n_{\uparrow\sigma l}(\textbf{r}) n_{\downarrow \sigma l}(\textbf{r}) \nonumber\\
+&V^{ll}_{1}[n_{Bl}(\textbf{r}+\boldsymbol{\delta}_l)+n_{Bl}(\textbf{r}+\hat{\textbf{x}}+\boldsymbol{\delta}_l)]n_{Al}(\textbf{r})\nonumber\\
+&V^{dp}_{1}[n_{B\bar{l}}(\textbf{r})+n_{B\bar{l}}(\textbf{r}+\hat{\textbf{x}})] n_{Al}(\textbf{r})\nonumber\\
+&V^{ll}_{2} n_{\sigma l}(\textbf{r})n_{\sigma l}(\textbf{r}+\hat{\textbf{x}})\nonumber\\
+&V^{dp}_{2}[n_{Ap}(\textbf{r})+n_{Ap}(\textbf{r}+\hat{\textbf{y}})]n_{Ad}(\textbf{r})\nonumber\\
+&V^{dp}_{2}[n_{Bd}(\textbf{r})+n_{Bd}(\textbf{r}+\hat{\textbf{y}})]n_{Bp}(\textbf{r}), 
\end{align}
where $\sigma$ and $l$ indices are summed over, $n_{s\sigma l}(\textbf{r})$ is the density with spin $s$ and orbital $l$ locating at sublattice $\sigma$ in the unit cell centered at $\textbf{r}$, $n_{\sigma l}(\textbf{r})= \sum_s n_{s\sigma l}(\textbf{r})$, and $\boldsymbol{\delta}_{p}=0$, $\boldsymbol{\delta}_{d}= \hat{y}$. 
As mentioned in the main text, $U^{l}$ denotes the on-site interactions for orbital $l$, and $V_1^{l\tilde{l}}$ ($V_2^{l\tilde{l}}$) denotes the nearest-neighbor (next nearest-neighbor) interactions on the zigzag chains with intra- or inter-orbital characters for $\tilde{l}=l$ and $\tilde{l}=\bar{l}$, respectively.

\section{II.~~~Competition between even- and odd-parity paired states}
Given the understanding about the real space configurations discussed in the main text, we now attempt to understand the phase diagram in Fig. 2(b) in the main text within a simplified ``two-patch'' scheme, where we ignore the intra-pocket momentum-dependence and consider effectively two points located at $\textbf{k}=(\pm k_F,0)$ instead of two pockets. Such simplification works well in small-pocket limit, and is similar in spirit to the Eliashberg formalism, where the momentum-dependence is assumed to be uniform within a pocket. 
Here we consider only the interactions responsible for the most dominant components in the self-consistency solutions with $A_g$ and $B_u$ symmetries [see the bonds with $\Delta_1$ in Fig.~2(c) and (d) in the main text], i.e. $U^d$ and $V_2^{dd}$ terms in Eq. 2 in the main text.  
Such interaction has a simple momentum dependence of $\mathbb{V}(\textbf{q})=U+V\cos(q_x)$, where we set $U^d=U$, $V_2^{dd}=V$, and $\textbf{q}$ the momentum transfer. 
Thus the interaction matrix $\Gamma$ in the linearized gap equation on p.3 in the main text can be simplified into a $2\times 2$ matrix in the basis of incoming and outgoing momenta running over $\textbf{k}^{(')}=(\pm k_F,0)$:
\begin{align}
\Gamma= \left(\begin{array}{cc}
\mathbb{V}(0) & \mathbb{V}(2k_F) \\
\mathbb{V}(-2k_F) & \mathbb{V}(0)
\end{array}\right).
\end{align}
The two eigenvalues $\frac{\mathbb{V}(0)\pm \mathbb{V}(2 k_F)}{2}$ correspond to the parity-even and odd eigenvectors $(1,1)$ and $(1,-1)$ respectively, and the eigenvector with the more negative eigenvalue corresponds to the dominant pairing gap. 
Evidently, repulsive $V$ forbids the parity-odd $A_u/B_u$ whereas repulsive $U$ forbids the parity-even $A_g$, given that $\cos(2k_F)<0$. This is true for a wide range of $k_F$ values, including that from the model in Eq. 1 in the main text. 
As for the cases where $U$ and $V$ are both attractive, the balance between $A_g$ and $B_u$ is tilted by the sign of the interaction with large momentum transfer $\mathbb{V}(2k_F)$. 
To be precise, when the $2k_F$ component contributed by $V$ dominates over the momentum-independent $U$ ($\mathbb{V}(2k_F)=U+V\cos(2k_F)>0$), the odd-parity $B_u$ is favored over the even-parity $A_g$. For the lightly gated WTe$_2$ model $H_0$ with $\mu=0.5$, the two Fermi pockets are located at $\textbf{k}=(\pm k_F,0)$ with $k_F\sim 1.2$. 

\section{III.~~~Majorana corner modes in various spin-triplet phases}

\begin{figure}[!]
\includegraphics[width=8cm]{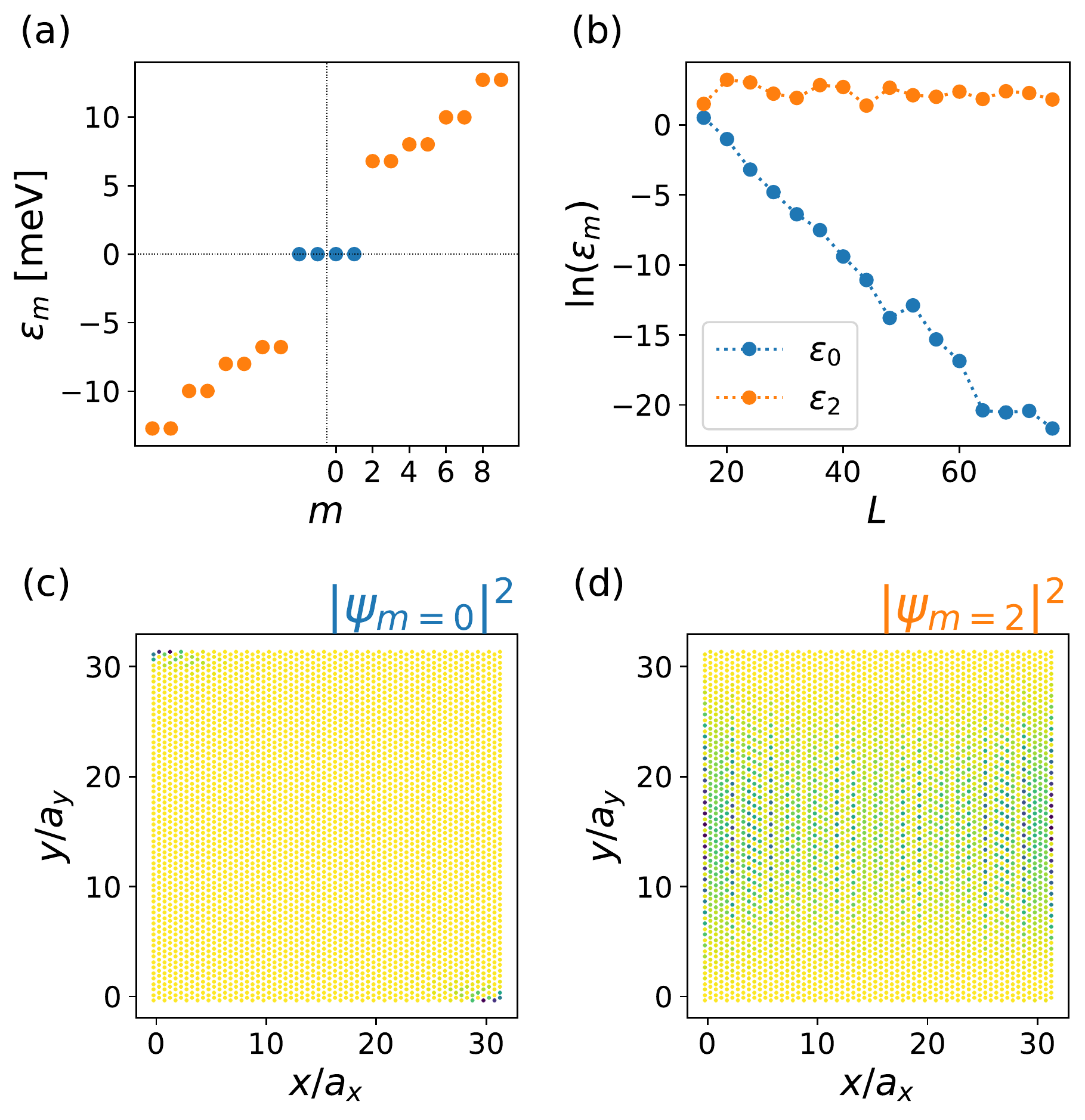}
\caption{More details about Fig. 3 in the main text: 
$B_u$ pairing symmetry at $U=-0.2$, $V=-0.4$ on a finite lattice of $L \times L$ unit cells. (a) and (c) are duplicated from the main text, while (b) shows the exponential scaling of the energy of the Majorana-Kramers' doublet with $L$, as well as the saturation of the spectral gap. (d) Spatial probability distribution of the lowest gapped state, which is a bulk state. For numerical convenience, here we take the self-consistent $B_u$ symmetry solution and multiply by 10, so that the resulting superconducting gaps are always much larger than any finite-size gaps of the normal bulk or edge states for tractable lattice sizes.}  
\label{MajoBu}
\end{figure}
\begin{figure}[!]
\includegraphics[width=8cm]{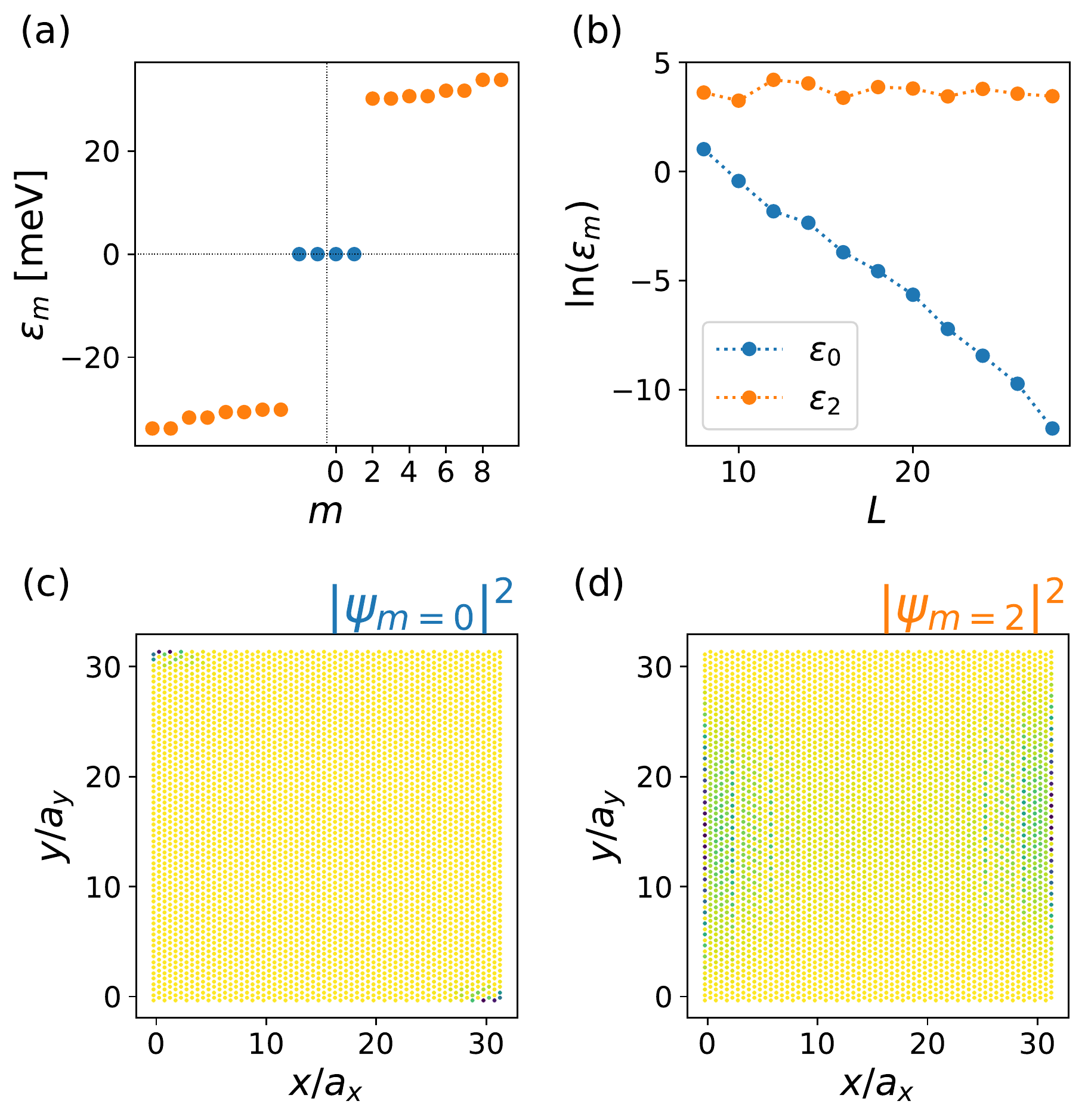}
\caption{Numerical evidences for corner Majorana Kramers' pairs in another spin-triplet phase $A_u$ in our phase diagram (see Fig. 2b in the main text). Here we put the self-consistent solution at $U=0.4$, $V=-1.0$, which has pairing symmetry $A_u$, on a finite lattice of $L \times L$ unit cells.
(a) The gapped spectrum with a zero energy Majorana-Kramers' doublet.
(b) The scaling of the corresponding eigenenergies for the $m=0$ state (the Majorana doublet) and the $m=2$ state (the lowest-energy gapped states) with increasing $L$.
(c) The spatial probability distribution $|\psi_0|^2$ corresponding to the zero modes, demonstrating sharp corner localization.
(d) Spatial probability distribution $|\psi_2|^2$ of the lowest gapped state.}  
\label{MajoAu}
\end{figure}
\begin{figure}[!]
\includegraphics[width=8cm]{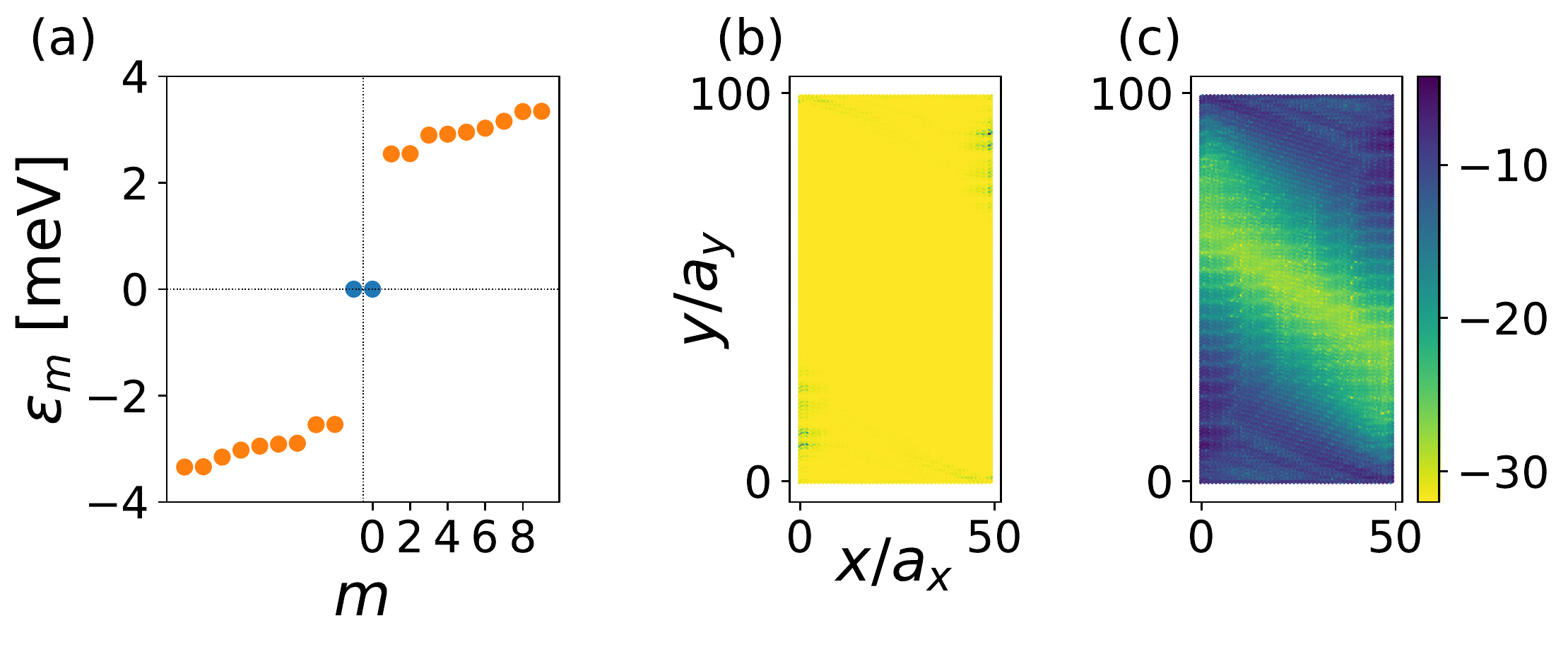}
\caption{More details about Fig. 4 in the main text: BdG spectrum for gated WTe$_2$ with the field-induced equal-spin superconducting phase $B_u''$ at $U=-1.0$, $V=-0.4$, $h_x/H_p\sim 5.3$ on a finite lattice of $50 \times 100$ unit cells.
    (a) The gapped spectrum with a single, near-zero energy Majorana mode. This is the Fig. 4b in the main text. 
    (b) The spatial probability distribution $|\psi_0|^2$ corresponding to the zero mode (this is the Fig. 4c in the main text), and
    (c) $\ln|\psi_0|^2$ which shows that the zero mode is exponentially localized near the corners. Here we take the self-consistent $B_u''$ symmetry solution and multiply by 10 for numerical convenience.}  
\label{MajoBupp}
\end{figure}
\begin{figure}[!]
\includegraphics[width=8cm]{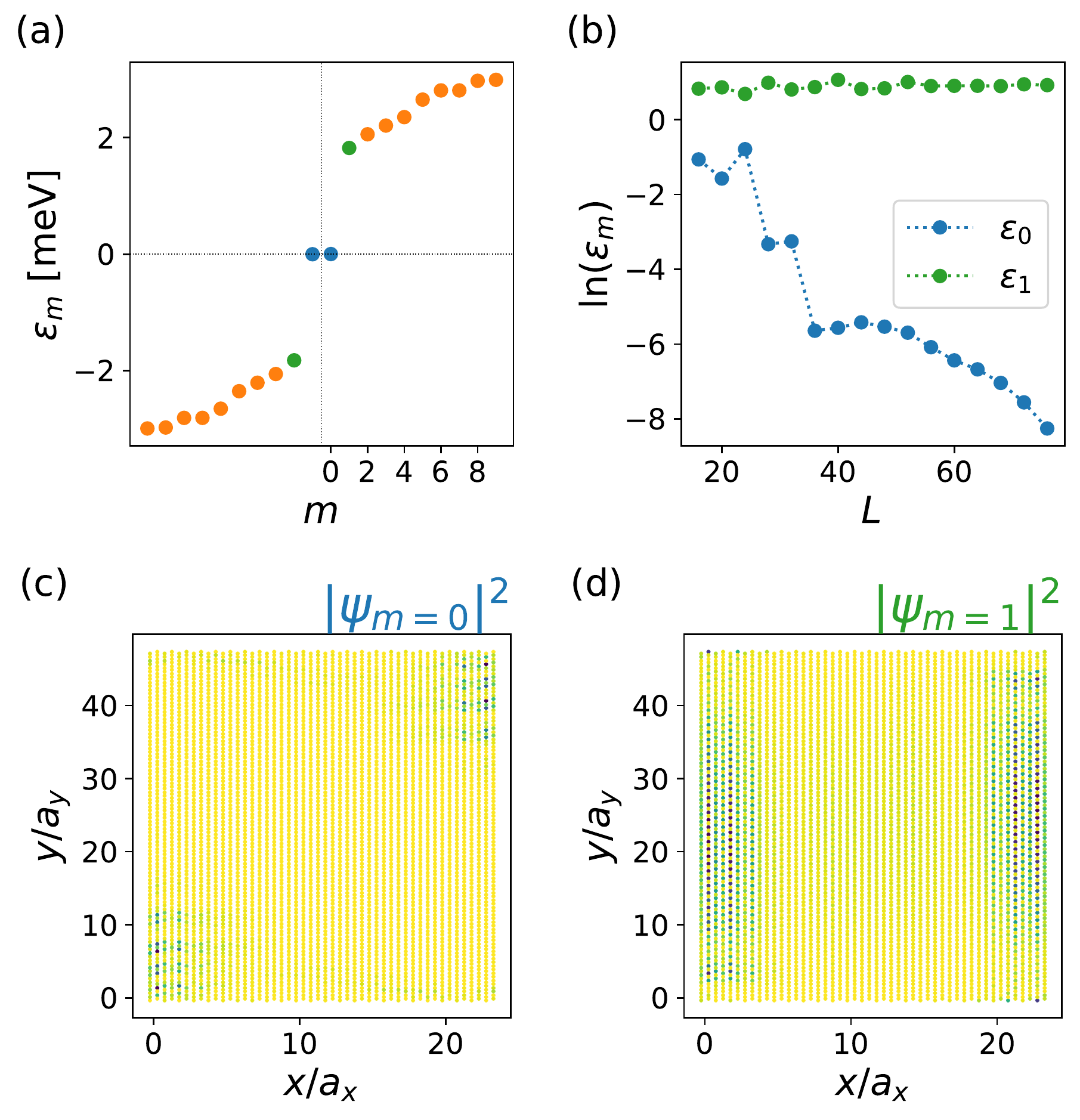}
\caption{
More numerical evidences for the existence of corner Majorana modes in the field-induced equal-spin phase $Bu''$ on a geometry smaller than that in Fig. \ref{MajoBupp}.    
The parameters used are the same as Fig. \ref{MajoBupp} except that the lattice is of $28 \times 48$ unit cells. Note that in the absense of time-reversal symmetry, there is only a single Majorana mode. Captions for the subfigures are otherwise the same as Fig.~\ref{MajoAu}. Here we also take the self-consistent $B_u''$ symmetry solution and multiply by 10 for numerical convenience.}  
\label{MajoBupp2}
\end{figure}
In this section, we show more numerical evidences for Majorana corner modes in different spin-triplet phases we find in WTe$_2$ phase diagram, the field-free phases $B_u$ and $A_u$, and the field-induced equal-spin phase $Bu^{''}$, as examples for our recipe (as these phases all clearly satisfy our recipe).  

We first show that the corner-localized Kramers' doublets we find for the $B_u$ phase [see Fig. 3 in the main text, also duplicated here in Fig. \ref{MajoBu}(a)(c)] indeed have zero energies by studying the finite-size scaling of the low-lying eigenvalues of the BdG Hamiltonian. 
Since we are interested only in the spectrum near zero, here we can partially diagonalize $H^{\rm BdG}$ using Lanczos techniques on extremely large lattices with open boundary conditions along $\hat{x}$ and $\hat{y}$ directions. 
It is clear from Fig. \ref{MajoBu}(b) that with an increasing lateral dimension $L$ of lattices with open boundary conditions along $\hat{x}$ and $\hat{y}$ directions, the corner-localized bound states tend exponentially toward zero energy with an increasing gap to the lowest-lying bulk quasiparticle excitations. 
In Fig. \ref{MajoBu}(d), we show the probability distribution of the lowest supragap state $|\psi_2|^2$ extends over the bulk. This is in contrast to the lowest eigenstate $|\psi_0|^2$ we showed in Fig. 3(b) in the main text [duplicated here in Fig. \ref{MajoBu}(c)], which sharply localized to the upper-left and bottom-right corners.

We now turn to the other spin-triplet phase $A_u$ in the superconducting phase diagram for WTe$_2$ in Fig. 2b in the main text. Since the normal state is a gated QSH material with Fermi pockets all away from TRIMs [see Fig. 2a in the main text] and the pairing is spin-triplet, phase $A_u$ satisfies our recipe and should host Majorana Kramers' pairs localized on opposite corners, just as phase $B_u$. 
Indeed, in Fig. \ref{MajoAu}(a) we show that the BdG spectrum indeed exhibits two zero-energy Kramers' pairs, and the finite-size scaling of low-lying eigen-energies in Fig. \ref{MajoAu}(b) further confirms that the doublets' energy $\epsilon_0\rightarrow 0$. 
Moreover, just as the $B_u$ case, the density distribution of these zero-energy states $|\psi_0|^2$ are sharply localized to two opposite corners, and that of the lowest supragap state $|\psi_2|^2$ extends over the bulk [see Fig. \ref{MajoAu}(c) and (d) respectively].

As we discussed in the main text, an in-plane magnetic field can drive a phase transition from the spin-singlet phase $A_g$ to the equal-spin phase $Bu_{''}$ [see Fig. 4 in the main text]. 
Here, we further show the log-scale density profile for the single Majorana modes we find in phase $Bu_{''}$ [see Fig. \ref{MajoBupp}(c)] to demonstrate their localization properties. For a localized state $\psi$, we expect $\left| \psi(\vec{r}) \right|^2 \leq C \exp \left( \left| \vec{r} - \vec{r}_0 \right| / \xi \right)$, where $\vec{r}_0$ is the localization center, $\xi$ a characteristic localization length, and $C$ some constant. Therefore, from Fig. \ref{MajoBupp}(c) we can better see that the wavefunction of a Majorana mode is exponentially localized to the two opposite corners from the roughly linear change in the color scale. 

Finally, in Fig. \ref{MajoBupp2} we present another set of results for the Majorana corner modes in phase $Bu_{''}$ on a rectangular geometry with different dimensions to show the robustness of our findings. 
In contrast to the time-reversal symmetric phases $A_u$ and $B_u$, the time-reversal broken BdG spectrum here shows two single zero-energy modes instead of two Kramers' doublets [see Fig. \ref{MajoBupp2}(a)]. In Fig. \ref{MajoBupp2}(b), we further show the finite-size scaling for the lowest two eigen-energies to confirm that the BdG spectrum is gapped and that Majorana modes indeed have zero energies. 
We then show that the Majorana modes are localized near two opposite corners [see Fig. \ref{MajoBupp2}(c)], and that the lowest gapped states are bulk states [see Fig. \ref{MajoBupp2}(d)]. It is interesting to find that the latter contains a remnant of the QSH edge states in the absense of pairing.

\section{IV.~~~Relating the Z$_4$ indicator and Z$_2$ index for 2D superconductors}
In this section, we explain how to obtain Eq. 5 in the main text. Consider a time-reversal and inversion-symmetric superconductor in 2D. Due to the presence of inversion symmetry, the $Z_2$ topological index $\nu$ for 2D time-reversal superconductors (class DIII) can be written in the following way: 
\begin{align}
(-1)^{\nu}=\prod_{\textbf{k}\in\rm{TRIM}}\prod_{m=1}^{n_{\textbf{k}}/2}\xi_{\textbf{k}2m},
\label{nudf}
\end{align}
where $\xi_{\textbf{k}m}$ is the parity of the filled BdG band $m$ at a time-reversal invariant momentum (TRIM). Here, the parity is the eigenvalue of a BdG band for the BdG inversion operator $I$ defined below Eq. 4 in the main text, $n_{\textbf{k}}$ is the total number of filled bands at $\textbf{k}$, and $n_{\textbf{k}}=n$ is independent of $\textbf{k}$ for superconductors. Only half of the bands enter the product since the system is two-fold degenerate. When $(-1)^{\nu}=1$ (0), the system is topological (trivial). Similar to the normal state case studied previously\cite{Indicator_PRX}, we can also express this index directly as 
\begin{align}
\nu=\sum_{\textbf{k}\in\rm{TRIM}}\frac{n^-_{\textbf{k}}}{2}, 
\label{nudf}
\end{align}
where $n^{\pm}_{\textbf{k}}$ is the number of occupied BdG bands with parity $\pm 1$. 

For cases where the normal state is half-filled, one can also define a Z$_4$ symmetry indicator $\kappa$ for 2D inversion-symmetric superconductors (in the presence of time-reversal symmetry) the way we do in Eq. 4 in the main text. To relate $\kappa$ to $\nu$, we further write it as 
\begin{align}
\kappa=\frac{1}{4}\sum_{\textbf{k}\in\rm{TRIM}}\sum_{m=1}^{n}\xi_{\textbf{k}m}
=\sum_{\textbf{k}\in\rm{TRIM}}\frac{n}{4}-\sum_{\textbf{k}\in\rm{TRIM}}\frac{n^-_{\textbf{k}}}{2}. 
\label{kappadf}
\end{align}
Here we have used the fact that $\sum_{m=1}^{n}\xi_{\textbf{k}m}=n^+_{\textbf{k}}-n^-_{\textbf{k}}=n-2n^-_{\textbf{k}}$. Now it is clear that 
\begin{align}
(-1)^{\nu}=(-1)^{\kappa} 
\label{nukappa}
\end{align}
since there are four TRIMs in 2D and that $n$ is even due to the two-fold degeneracy. We therefore conclude that 
\begin{align}
\nu=\kappa~~\rm mod~~2,   
\label{nukappa}
\end{align}
which is the Eq. 5 in the main text.

\section{V.~~~Parity eigenvalues for BdG bands}
\begin{table}[]
\centering
\begin{tabular}{ |c|c|c|c|c|c|c|c|c| }
\hline
   $n$ & 1 & 2 & 3 & 4 & 5 & 6 & 7 & 8 \\
 \hline
 $\Gamma$ &- &- & + &+ &- &- &+ &+\\
 \hline
 $X$ &- &- & + &+ &- &- &+ &+\\
 \hline
 $Y$ &+ &+ &+ &+ &+ &+ &+ &+\\
 \hline
 $M$ &- &- & + &+ &- &- &+ &+\\
 \hline
\end{tabular}\\
\caption{
The inversion eigenvalues $p_{\tilde{\textbf{k}},n}$ of all the occupied $\rm BdG$ bands $n=1, \cdots, 8$, ordered with increasing energy, at each of the TRIMs $\tilde{\textbf{k}}=(0,0)$, $(\pi,0)$, $(0,\pi)$ and $(\pi,\pi)$. The indicator $\kappa$ for inversion-protected topological superconductors is given by the sum of these eigenvalues divided by 4.
}
\label{indicator}
\end{table}
To calculate the symmetry indicator $\kappa$ for the BdG Hamiltonian of monolayer WTe$_2$ with the odd-parity $B_u$ pairing, we need to calculate the parity eigenvalues of the filled BdG bands according to the definition of $\kappa$ [see Eq. 4 in the main text]. We numerically obtain these eigenvalues at each of the high-symmetry points for our model $H_0$ with the self-consistently obtained $B_u$ pairing, as shown in Table \ref{indicator}. 

\section{VI.~~~Relating the Z$_2$ indicator and Z$_2$ index for topological `metals'}
In this section, we explain why the second equation in Eq. 5 in the main text holds for topological `metals' that are `effectively gapped'. We first review the insulator case presented in Ref. \onlinecite{Indicator_PRX}. Consider an insulator with both time-reversal and inversion symmetries. Very similar to the DIII superconductor case in section IV., the Z$_2$ index $\nu_N$ for time-reversal topological insulators can be calculated simply by\cite{FuKaneZ2} 
\begin{align}
(-1)^{\nu_N}=\prod_{\textbf{k}\in\rm{TRIM}}\prod_{m=1}^{n_{\textbf{k}}/2}\xi_{\textbf{k}2m},
\label{nuNdf}
\end{align}
where $\xi_{\textbf{k}m}$ is the parity of the filled normal band $m$ at a time-reversal invariant momentum (TRIM). Here, the parity is the eigenvalue of a normal band for the `normal' inversion operator $I_0$ defined in the main text, $n_{\textbf{k}}$ is the total number of filled normal bands at $\textbf{k}$, and  $n_{\textbf{k}}=n$ is independent of $\textbf{k}$ for insulators. Only half of the bands enter the product since the system is two-fold degenerate. When $(-1)^{\nu_N}=1$ (0), the system is topological (trivial). As shown in Ref. \onlinecite{Indicator_PRX}, we can also express this index directly as 
\begin{align}
\nu_N=\sum_{\textbf{k}\in\rm{TRIM}}\frac{n^-_{\textbf{k}}}{2}, 
\label{nuNdf}
\end{align}
where $n^{\pm}_{\textbf{k}}$ is the number of occupied normal bands with parity $\pm 1$. 

Due to the presence of inversion symmetry, one can also calculate the Z$_2$ symmetry indicator $\kappa_N$ for inversion-protected topological crystalline insulators given by 
\begin{align}
\kappa_N=\frac{1}{4}\sum_{\textbf{k}\in\rm{TRIM}}\sum_{m=1}^{n_{\textbf{k}}}\xi_{\textbf{k}m}
=\sum_{\textbf{k}\in\rm{TRIM}}\frac{n_{\textbf{k}}}{4}-\sum_{\textbf{k}\in\rm{TRIM}}\frac{n^-_{\textbf{k}}}{2}. 
\label{kappaNdf}
\end{align}
Here we have used the fact that $\sum_{m=1}^{n_{\textbf{k}}}\xi_{\textbf{k}m}=n^+_{\textbf{k}}-n^-_{\textbf{k}}=n_{\textbf{k}}-2n^-_{\textbf{k}}$. Now it is clear that 
\begin{align}
(-1)^{\nu_N}=(-1)^{\kappa_N} 
\label{nukappa}
\end{align}
if $\alpha\equiv\sum_{\textbf{k}\in\rm{TRIM}}\frac{n_{\textbf{k}}}{4}$ is even. Since both indices are $Z_2$, we have $\nu_N=\kappa_N$.

For insulators, $\alpha=n$(number of TRIMs)$/4$, where $n$ is even due to the two-fold degeneracy. Eq. \ref{nukappa} in the Supplementary Material is thus not guaranteed for 1D insulators, but holds for both 3D and 2D insulators, where the numbers of TRIMs are 8 and 4, respectively. 
As for metals with Fermi pockets enclosing any TRIM, $n_{\textbf{k}}\neq n$ is different for different TRIMs in general, and thus $\alpha$ is not guaranteed even. 
Nonetheless, for metals whose Fermi pockets are away from TRIMs, $n_{\textbf{k}}=n$ is still $\textbf{k}$-independent, and thus Eq. \ref{nukappa} in the Supplementary Material still holds in 2D and 3D.\\

\section{VII.~~~Bulk-boundary correspondence for $\kappa=2$ inversion-protected TCsc}
As mentioned in the main text, the fact that a $\kappa=2$ inversion-protected TCsc in 2D has no Majorana edge modes but is still topologically distinct from the trivial $\kappa=0$ phase suggests the possibility of Majorana corner modes. 
Such bulk-boundary correspondence for a higher-order topological superconductor, however, has not been rigorously proven to the best of our knowledge. 
Thus to gain more understanding about this bulk-boundary correspondence, in this section we will first present a minimal model for a $\kappa=2$ TCsc in 2D, which can be tuned across phase boundaries to $\kappa=3$ and 4 (trivial) phases. Then within this model, we will show analytically how zero-dimensional Majorana modes arise on the boundary of the $\kappa=2$ phase when placed against the trivial phase. 

\subsection{A.~~~Minimal model for an inversion-protected TCsc}
Our eight-band minimal model $\mathcal{H}=\sum_{\textbf{k}}\mathcal{H}(\textbf{k})$ 
\begin{align}
\mathcal{H}(\textbf{k}) &= \epsilon_0\hat{\tau}_z\otimes\hat{s}_0\otimes\hat{\rho}_0\nonumber\\
&+[m_0+m_1(\cos k_x+\cos k_y)]\hat{\tau}_z\otimes\hat{s}_0\otimes\hat{\rho}_z\nonumber\\
&+v\sin k_x\hat{\tau}_0\otimes\hat{s}_z\otimes\hat{\rho}_x+v\sin k_y\hat{\tau}_z\otimes\hat{s}_0\otimes\hat{\rho}_y\nonumber\\
&+\Delta\sin k_x\hat{\tau}_x\otimes\hat{s}_z\otimes\hat{\rho}_0+\Delta\sin k_y\hat{\tau}_y\otimes\hat{s}_0\otimes\hat{\rho}_z    
\label{Hlat}
\end{align}
consists of a regularized Bernevig-Hughes-Zhang (BHZ) like model for a QSH state with odd-parity pairing.  
Here $\hat{\tau}$, $\hat{s}$, and $\hat{\rho}$ are Pauli matrices for particle and hole, spin $s=\uparrow,\downarrow$, and orbital $\rho=s,p_-$. 
This model $\mathcal{H}$ is invariant under time reversal operation $\Theta=is_y\mathcal{K}$, $\textbf{k}\rightarrow-\textbf{k}$, particle-hole transformation $\mathcal{P}=\hat{\tau}_x\mathcal{K}$, $\textbf{k}\rightarrow-\textbf{k}$, and the inversion operation defined for odd-parity superconductors $\textit{I}$ $=\hat{\tau}_z\otimes\hat{\rho}_z$,  $\textbf{k}\rightarrow-\textbf{k}$. 

This minimal model exhibits two bulk topological transitions in the parameter space of $\epsilon_0$, $m_0$, and $m_1$: one is from the inversion symmetry indicator $\kappa=2$ to $\kappa=3$, and the other from $\kappa=3$ to $\kappa=4$. 
In particular, we can tune through different phases by tuning $m_1$ at a fixed $m_0$ and $\epsilon_0\geq 0$ as follows 
\begin{align}
&\kappa=4 (0):~~~-(m_0-\epsilon_0)<2m_1<m_0-\epsilon_0\nonumber\\
&\kappa=3:~~~~~~~-(m_0+\epsilon_0)<2m_1<-(m_0-\epsilon_0)\nonumber\\
&\kappa=2:~~~~~~~~~~~~~~~~~~~~~~~~~~~2m_1<-(m_0+\epsilon_0),   
\label{PT}
\end{align}
where all band inversions occur at $\Gamma$. 
Taking the topologically trivial $\kappa=4$ phase as the reference point, the spectrum first undergoes a single band inversion to enter the $\kappa=3$ phase, then follows another band inversion to enter the $\kappa=2$ phase. 

\begin{figure}[!]
\includegraphics[width=8cm]{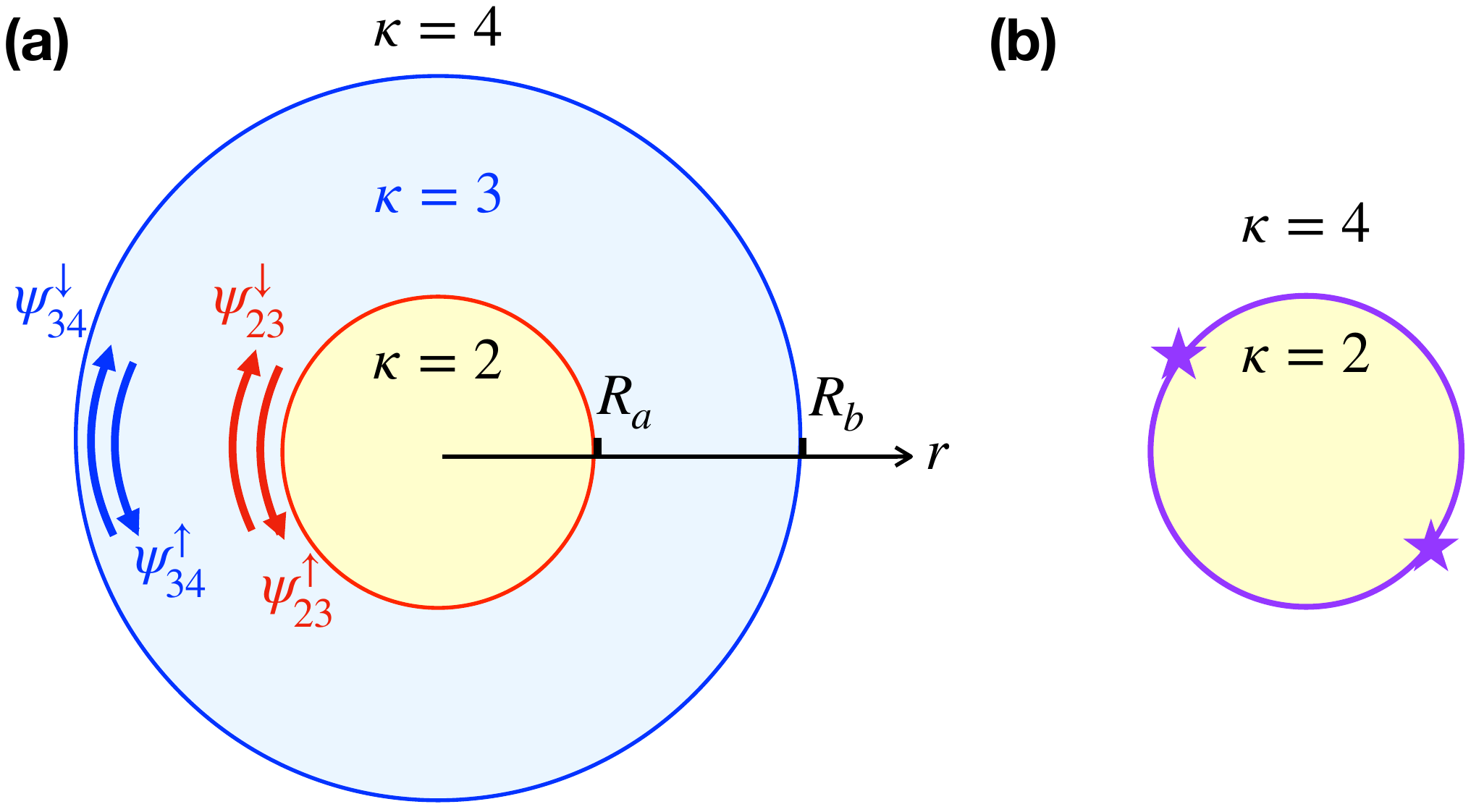}
\caption{Schematics for the Majorana boundary modes of the minimal model on (a) a three-domain geometry containing phases with $\kappa=2$, $3$, and $4$ (trivial), and (b) an open geometry for the $\kappa=2$ phase. In (a), the $\kappa=2$, 3, and 4 phases live in the inner, middle, and outer domains separated by domain walls at $r=R_a$ (red circle) and $R_b$ (blue circle). 
The red and blue arrows represent the helical edge modes $\psi_{23}^{s}$ and $\psi_{34}^{s'}$ of the $\kappa=3$ phase on the domain walls against the $\kappa=2$ and 4 phases, respectively. The geometry in (b) can be obtained from (a) by shrinking the $\kappa=3$ domain to zero. The purple circle represents the resulting boundary between the $\kappa=2$ and 4 phases, and the two purple stars represent the zero-dimensional Majorana modes that localize on this new boundary. These `higher-order' Majorana modes can be viewed as the `leftover' edge modes in (a) that survive the symmetry-allowed perturbations.}  
\label{eff}
\end{figure}

\subsection{B.~~~Edge modes in a three-domain geometry}
To understand what kind of boundary modes a $\kappa=2$ phase can host when placed against a trivial phase, we first study the edge modes in a geometry of concentric rings with three domains along the radial direction $r$: $\kappa=2$ phase for $r<R_a$, $\kappa=3$ phase for $R_a<r<R_b$, and $\kappa=4$ phase for $r>R_b$ [see Fig. \ref{eff} (a) in Supplementary]. Since $\kappa$ is defined modulo 4, the phase $\kappa=4\equiv 0$ on the outside is the trivial phase. 
Given that all band inversions in $\mathcal{H}$ happen at $\Gamma$, 
it is more convenient to realize this geometry by working with the $k\cdot p$ model $\mathcal{H}_{\Gamma}(\textbf{k})$ around $\Gamma$ written in the polar coordinate $(r,\theta)$. 
We focus on the $v\ll\Delta$ limit to avoid the possible existence of edge modes, in which case the existence of the submerged corner modes becomes ambiguous.  
Together with the fact that $s=\uparrow,\downarrow$ is a good quantum number, we arrive at the Hamiltonian for each spin species in the $\hat{\tau}\otimes\hat{\rho}$ basis
\begin{align}
&\mathcal{H}^{\uparrow/\downarrow}_\Gamma (r,\theta)\nonumber\\
&= \left(\begin{array}{cccc}
m_+(r) & 0 & \pm\Delta e^{\mp i\theta} k_r & 0 \\
0 & m_-(r) & 0 & \pm\Delta e^{\pm i\theta}k_r \\
\pm\Delta e^{\pm i\theta}k_r & 0 & -m_+(r) & 0 \\
0 & \pm\Delta e^{\mp i\theta}k_r & 0 & -m_-(r)  
\end{array}\right),
\label{hgamma}
\end{align}
where $k_r=-i\partial_r$ and $m_{\pm}(r)\equiv \epsilon_0(r)\pm(m_0(r)+2m_1(r))$. 
Here we have set $1/r\rightarrow 0$ since we focus on the asymptotic form of the edge modes (i.e. consider $|r-R_{a,b}|\gg 0$). 
The three-domain geometry can then be realized by setting
\begin{align}
&m_{+}(r)<0,~~m_{-}(r)>0 ~~\mbox{for}~~r<R_a\nonumber\\
&m_{+}(r)>0,~~m_{-}(r)>0 ~~\mbox{for}~~R_a<r<R_b\nonumber\\
&m_{+}(r)>0,~~m_{-}(r)<0 ~~\mbox{for}~~r>R_b. 
\label{mass1}
\end{align}

Since the $\kappa=3$ superconducting phase in the middle domain has a nontrivial $Z_2$ index $\nu=1$, we expect to find zero-energy eigenstates $\psi^s_{23}(r,\theta)$ and $\psi^{s'}_{34}(r,\theta)$ localizing along the domain walls at $r=R_a$ and $r=R_b$, respectively. 
By taking $m_+(r)=$sgn$(r-R_a)$, $m_-(r)=-$sgn$(r-R_b)$, and $\Delta>0$ for simplicity, we find one pair of helical edge modes per domain wall {\color{magenta}[see Supplementary Material Sec. IV A]}: 
\begin{align}
&\psi^{\uparrow/\downarrow}_{23}(r,\theta)=e^{-\frac{1}{\Delta}|r-R_a|}e^{il\theta}\left(\begin{array}{c}
e^{\mp i\frac{\theta}{2}} \\0 \\\pm ie^{\pm i\frac{\theta}{2}} \\0 \end{array}\right),
\label{psi23up}
\end{align}
and 
\begin{align}
&\psi^{\uparrow/\downarrow}_{34}(r,\theta)=e^{-\frac{1}{\Delta}|r-R_b|}e^{il\theta}\left(\begin{array}{c}
0 \\e^{\pm i\frac{\theta}{2}} \\0 \\\mp ie^{\mp i\frac{\theta}{2}} \end{array}\right),  
\label{psi34up}
\end{align}
where $l$ is the orbital angular momentum taking half integers. 
Given that $\psi^s_{23/34}(r,\theta)$ obeys Majorana condition up to an overall phase and that the spin-up modes $\psi^{\uparrow}_{23/34}$ and spin-down modes $\psi^{\downarrow}_{23/34}$ propagate along the domain walls in opposite directions {\color{magenta}[see Supplementary Material Sec. IV B]}, we have arrived at one pair of helical Majorana edge modes per domain wall [see Fig. \ref{eff} (a) in Supplementary Material]. 

\subsection{C.~~~Majorana corner modes in a $\kappa=2$ phase}
We are now ready to study the boundary modes between a $\kappa=2$ and a $\kappa=4$ phase. In the following we will shrink the $\kappa=3$ domain by bringing $R_a\rightarrow R_b$, and see if there exists any symmetry-allowed perturbation that gaps out the edge modes $\psi^s_{23}$ and $\psi^{s'}_{34}$ at the two domain walls.  

We first write down the rotational invariant perturbations $\mathcal{H}'_{\rm rot}(r,\theta)$ up to linear oder, and project them onto the edge modes from the two domain walls {\color{magenta}[see Supplementary Material Sec. IV C]}. We find that $\mathcal{H}'_{\rm rot}(r,\theta)$ only couples edge modes propagating in the same direction, i.e. $\psi^{s}_{23}$ and $\psi^{s}_{34}$. The two pairs of helical edge modes therefore remain gapless in the presence of rotational invariant perturbations.


Next we consider the rotational-breaking perturbations. Here we focus on the lowest order terms $\mathcal{H}'$, which have no spatial dependence. After projecting all the symmetry-allowed terms onto the edge modes $\psi_{23}(r,\theta)$ and $\psi_{34}(r,\theta)$, we find that there are only two non-vanishing terms that are hermitian and couple components in the edge modes that are counter-propagating 
\begin{align}
&\mathcal{H}_{a}^{'}=\hat{\tau}_x\otimes \hat{s}_x\otimes\hat{\rho}_y\nonumber\\
&\mathcal{H}_{b}^{'}=\hat{\tau}_y\otimes \hat{s}_y\otimes\hat{\rho}_x. 
\label{pert0}
\end{align}
Both of these terms are pairing terms that are odd under inversion, i.e. $[\mathcal{H}',\textit{I}~]=0$, where \textit{I} is the inversion operator for odd-parity superconductors defined earlier.  
Moreover, their corresponding amplitudes after projection $\tilde{\mathcal{H}}'_{a,b}(\theta)=\int dr \psi_{23}^{\dagger}(r,\theta)\mathcal{H}'_{a,b}\psi_{34}(r,\theta)$ have an angular dependence of $\tilde{\mathcal{H}}'_{a}(\theta)\propto i\sin\theta$ and $\tilde{\mathcal{H}}'_{b}(\theta)\propto i\cos\theta$, respectively. 
The full Hamiltonian of the edge modes therefore has the form 
\begin{align}
\tilde{\mathcal{H}}'_{\rm edge}(\theta)
=\frac{2l}{R_b}\hat{\lambda}_0\hat{s}_z+(\alpha\cos\theta+\beta\sin\theta)\hat{\lambda}_y\hat{s}_x, 
\label{proj0}
\end{align}
where $\alpha$ and $\beta$ are real numbers, and $\hat{\lambda}$, $\hat{s}$ are Pauli matrices for edge modes $\psi^s_{23}$, $\psi^s_{34}$, and spin $s=\uparrow,\downarrow$, respectively. 
Since the back-scattering term $\alpha\cos\theta+\beta\sin\theta$ has opposite signs at any $\theta$ and $\theta+\pi$, it is bound to vanish at $\theta_0=\tan^{-1}(\frac{\beta}{\alpha})+\frac{\pi}{2}$ and $\theta_0+\pi$.
This means that even at the lowest order, rotational breaking perturbations will gap the edge modes $\psi_{23}$ and $\psi_{34}$ in an odd-parity way, 
and a $\kappa=2$ TCsc will host at least two `leftover' zero-dimensional zero-energy Kramer's pairs located at $\theta_0$ and $\theta_0+\pi$ in an open geometry [see Fig. \ref{eff} (b) in Supplementary Material]. The specific value of $\theta_0$ is given by the microscopics, and these two zero-dimensional Majorana Kramers pairs are often trapped at the opposite corners of the considered geometry. 
Importantly, the two Majorana pairs can annihilate each other only when inversion symmetry is broken.
Thus within our minimal model for an inversion-protected TCsc with $\kappa=2$, we have shown how zero-dimensional Majorana Kramers pairs emerge on the boundaries, and hence an inversion-protected higher-order topological superconductor. 

\section{VIII.~~~Details regarding derivations in section V}
\subsection{A.~~~Edge modes in the three-domain geometry}
In this subsection, we show how we obtain the Majorana edge modes $\psi^{s}_{23}(r,\theta)$ and $\psi^{s'}_{34}(r,\theta)$ in the three-domain geometry. Consider the domain wall between the $\kappa=2$ and $\kappa=3$ phases. 
The Hailtonian $\mathcal{H}_{\Gamma}^{s}$ in Eq. \ref{hgamma} in Supplementary Material section III. becomes block diagonal after taking the limit $v\rightarrow 0$, and it is clear from Eq. \ref{hgamma} in Supplementary Material section III. that only the block with $m_+(r)$ terms experiences a sign-changing mass term and is thus expected to trap a zero-energy eigenstate $\psi^{\uparrow}_{23}(r,\theta)$ localized at $r=R_a$. In the following, we solve for the asymptotic for m of $\psi^{\uparrow}_{23}(r,\theta)$. This $2\times 2$ block (represented by Pauli matrix $\tau$) with $m_+(r)$ term for the spin-up sector is given by 
\begin{align}
h_+^{\uparrow}&=m_+(r)\hat{\tau}_z+\Delta(-i\partial_r)(\cos\theta\hat{\tau}_x+\sin\theta\hat{\tau}_y)\nonumber\\
&=m_+(r)\hat{\tau}_z+\Delta(-i\partial_r)\hat{\tau}_xe^{i\hat{\tau}_z\theta}.
\label{proj0}
\end{align}
The zero-energy eigenstate is then given by $h_+^{\uparrow}\psi^{\uparrow}_{23}=0$, but 
we can obtain the edge state more conveniently by solving the zero-energy mode for a rotated Hamiltonian $\tilde{h}_+^{\uparrow}\phi=0$, where $\tilde{h}_+^{\uparrow}=\hat{\tau}_x\hat{U}_{\theta}h_+^{\uparrow}\hat{U}^{\dagger}_{\theta}$, and $\hat{U}_{\theta}=e^{i\hat{\tau}_z\theta/2}$ is a unitary transformation. After some algebra, we find that
\begin{align}
\tilde{h}_+^{\uparrow}(r)&=\hat{\tau}_x[m_+(r)\hat{\tau}_z+\Delta(-i\partial_r)\hat{\tau}_x]\nonumber\\
&=-im_+(r)\hat{\tau}_y+\Delta(-i\partial_r)\hat{\tau}_0. 
\label{proj0}
\end{align}
Now $\tilde{h}_+^{\uparrow}$ is $\theta$-independent, and we can thus write its zero-energy eigenstate as $\phi(r)=f(r)\xi$, where the spinor $\xi$ obeys $\hat{\tau}_y\xi=a\xi$. 
For simplicity, we take $m_+(r)=$sgn$(r-R_a)$ and $\Delta>0$. 
Then by solving the differential equation for $f(r)$ and requiring that $f(r)$ localized at the boundary $r=R_a$, we find that $a=1$ and $\phi(r)=e^{-1/\Delta|r-R_a|}(1,0,i,0)^T$. Together with the fact that the orbital part of the angular momentum has the form $L_z=-i\partial_{\theta}$, we can then obtain the zero-mode for $h_+^{\uparrow}(r,\theta)$ with angular dependence 
\begin{align}
&\psi^{\uparrow}_{23}(r,\theta)=U^{\dagger}_{\theta}\phi(r)
=e^{-\frac{1}{\Delta}|r-R_a|}e^{il\theta}\left(\begin{array}{c}
e^{-i\frac{\theta}{2}} \\0 \\e^{i\frac{\theta}{2}} \\0 \end{array}\right).
\label{psi23up}
\end{align}
Note that the Majorana condition $\Xi\psi^{\uparrow}_{23}=\psi^{\uparrow}_{23}$ is satisfied up to an overall phase. $\psi^{\downarrow}_{23}(r,\theta)$ and $\psi^{\uparrow/\downarrow}_{34}(r,\theta)$ can be obtained in a similar way.  

\subsection{B.~~~Propagating directions of the edge modes}
To determine the propagating directions of these edge modes, we need to take back the terms containing $k_{\theta}=-i\partial_{\theta}$ in the $k\cdot p$ Hamiltonian $\mathcal{H}^{\uparrow/\downarrow}_\Gamma (\textbf{k})$ and consider them as perturbations. These are the terms that emerge when we write $k_{\pm}=k_r+ik_{\theta}$. In the Hamiltonian $\mathcal{H}^{\uparrow/\downarrow}_\Gamma (r,\theta)$ 
in Eq. \ref{hgamma} in Supplementary Material, we did not include these terms since we expect them to be small in the limit of large $r$. 
Such perturbation has the form 
\begin{widetext}
\begin{align}
&\mathcal{H}_{\theta}^{\uparrow/\downarrow}(r,\theta)= \left(\begin{array}{cccc}
0 & 0 & -i\Delta \frac{e^{\mp i\theta}}{r} k_\theta & 0 \\
0 & 0 & 0 & i\Delta \frac{e^{\pm i\theta}}{r} k_\theta \\
i\Delta \frac{e^{\pm i\theta}}{r} k_\theta & 0 & 0 & 0 \\
0 & -i\Delta \frac{e^{\mp i\theta}}{r} k_\theta & 0 & 0  
\end{array}\right). 
\label{htheta}
\end{align}
\end{widetext}
Recall that the edge modes are given by 
\begin{align}
&\psi^{\uparrow/\downarrow}_{23}(r,\theta)=e^{-\frac{1}{\Delta}|r-R_a|}e^{il\theta}\left(\begin{array}{c}
e^{\mp i\frac{\theta}{2}} \\0 \\\pm ie^{\pm i\frac{\theta}{2}} \\0 \end{array}\right),
\label{psi23up}
\end{align}
and 
\begin{align}
&\psi^{\uparrow/\downarrow}_{34}(r,\theta)=e^{-\frac{1}{\Delta}|r-R_b|}e^{il\theta}\left(\begin{array}{c}
0 \\e^{\pm i\frac{\theta}{2}} \\0 \\\mp ie^{\mp i\frac{\theta}{2}} \end{array}\right),  
\label{psi34up}
\end{align}
where $l$ is the orbital angular momentum taking half integers. 
We thus find the energy correction arising from rotational motion in $\theta$ to be 
\begin{align}
&\langle \mathcal{H}_{\theta}^{\uparrow}\rangle_{j,\uparrow}=(\psi^{\uparrow}_{j})^{\dagger}\mathcal{H}_{\theta}^{\uparrow}\psi^{\uparrow}_{j}\propto 2l\nonumber\\
&\langle \mathcal{H}_{\theta}^{\downarrow}\rangle_{j,\downarrow}=(\psi^{\downarrow}_{j})^{\dagger}\mathcal{H}_{\theta}^{\downarrow}\psi^{\downarrow}_{j}\propto -2l, 
\label{direction}
\end{align}
where $j=23,34$ denotes the two domain walls. We therefore conclude that the spin-up and spin-down modes are right- and left-movers respectively for both domain walls, which amounts to one pair of helical edge modes per domain wall [see Fig. \ref{eff} (a) in Supplementary Material]. \\

\subsection{C.~~~Rotational invariant perturbations}
In this subsection, we write down the general form of rotational invariant perturbations $H'_{\rm rot}(\textbf{k})$ and its projection on to the edge modes $\psi^{s}_{23}$ and $\psi^{s'}_{34}$. 
The rotational operator is given by $C_{\theta}=e^{-iJ_z\theta/2}$, $k_{\pm}\rightarrow e^{\mp i\theta}k_{\pm}$, where the angular momentum has the form $J_z=\hat{\tau}_z\otimes \hat{s_z}\otimes\hat{\sigma_z}/2$, and $k_{\pm}=k_x\pm ik_y$.   
The perturbations that preserve time-reversal symmetry $\Theta$, particle-hole symmetry $\mathcal{P}$, inversion symmetry for odd-parity pairing \textit{I}, and also obey $C_\theta H'_{\rm rot}(\textbf{k})C_\theta^{\dagger}=H'_{\rm rot}(\textbf{k})$ have the general form of 
\begin{widetext}
\begin{align}
&H'_{\rm rot}(\textbf{k})= \left(\begin{array}{cccccccc}
m_1 & Ak_- & 0 & 0 & Bk_- & D_0 & 0 & 0 \\
A^*k_+ & m_2 & 0 & 0 & -D_0^* & Ck_+ & 0 & 0 \\
0 & 0 & m_1 & -A^*k_+ & 0 & 0 & -B^*k_+ & D_0^* \\
0 & 0 & -Ak_- & m_2 & 0 & 0 & -D_0 & -C^*k_- \\
B^*k_+ & -D_0 & 0 & 0 & -m_1 & A^*k_+ & 0 & 0 \\
D_0^* & C^*k_- & 0 & 0 & Ak_- & -m_2 & 0 & 0 \\
0 & 0 & -Bk_- & -D_0^* & 0 & 0 & -m_1 & -Ak_- \\
0 & 0 & D_0 & -Ck_+ & 0 & 0 & -A^*k_+ & -m_2 \\
\end{array}\right).  
\label{Hrot}
\end{align}
\end{widetext}
Here, $H'_{\rm rot}(\textbf{k})$ is written in the basis of $\hat{\tau}\otimes \hat{s}\otimes\hat{\rho}$, and $m_{1/2}$, $D_0$, $A$, $B$, and $C$ are free parameters.  

Since the edge modes with the same spin (opposite spins) $\psi^s_{23}$ and $\psi^s_{34}$ ($\psi^{\bar{s}}_{34}$) propagate in the same direction (opposite directions), $H'_{\rm rot}$ has to couple edge modes with opposite spins in order to create a gap. However, it is obvious from Eq. \ref{Hrot} in Supplementary Material that there exists no spin-flipping terms that are allowed by the symmetries considered. The edge modes $\psi^s_{23}$ and $\psi^{s'}_{34}$ thus remain gapless in the presence of rotational invariant perturbations.


\begin{thebibliography}{63}%
\makeatletter
\providecommand \@ifxundefined [1]{%
 \@ifx{#1\undefined}
}%
\providecommand \@ifnum [1]{%
 \ifnum #1\expandafter \@firstoftwo
 \else \expandafter \@secondoftwo
 \fi
}%
\providecommand \@ifx [1]{%
 \ifx #1\expandafter \@firstoftwo
 \else \expandafter \@secondoftwo
 \fi
}%
\providecommand \natexlab [1]{#1}%
\providecommand \enquote  [1]{``#1''}%
\providecommand \bibnamefont  [1]{#1}%
\providecommand \bibfnamefont [1]{#1}%
\providecommand \citenamefont [1]{#1}%
\providecommand \href@noop [0]{\@secondoftwo}%
\providecommand \href [0]{\begingroup \@sanitize@url \@href}%
\providecommand \@href[1]{\@@startlink{#1}\@@href}%
\providecommand \@@href[1]{\endgroup#1\@@endlink}%
\providecommand \@sanitize@url [0]{\catcode `\\12\catcode `\$12\catcode
  `\&12\catcode `\#12\catcode `\^12\catcode `\_12\catcode `\%12\relax}%
\providecommand \@@startlink[1]{}%
\providecommand \@@endlink[0]{}%
\providecommand \url  [0]{\begingroup\@sanitize@url \@url }%
\providecommand \@url [1]{\endgroup\@href {#1}{\urlprefix }}%
\providecommand \urlprefix  [0]{URL }%
\providecommand \Eprint [0]{\href }%
\providecommand \doibase [0]{http://dx.doi.org/}%
\providecommand \selectlanguage [0]{\@gobble}%
\providecommand \bibinfo  [0]{\@secondoftwo}%
\providecommand \bibfield  [0]{\@secondoftwo}%
\providecommand \translation [1]{[#1]}%
\providecommand \BibitemOpen [0]{}%
\providecommand \bibitemStop [0]{}%
\providecommand \bibitemNoStop [0]{.\EOS\space}%
\providecommand \EOS [0]{\spacefactor3000\relax}%
\providecommand \BibitemShut  [1]{\csname bibitem#1\endcsname}%
\let\auto@bib@innerbib\@empty
\bibitem [{\citenamefont {Qian}\ \emph {et~al.}(2014)\citenamefont {Qian},
  \citenamefont {Liu}, \citenamefont {Fu},\ and\ \citenamefont
  {Li}}]{qian_2014sci}%
  \BibitemOpen
  \bibfield  {author} {\bibinfo {author} {\bibfnamefont {X.}~\bibnamefont
  {Qian}}, \bibinfo {author} {\bibfnamefont {J.}~\bibnamefont {Liu}}, \bibinfo
  {author} {\bibfnamefont {L.}~\bibnamefont {Fu}}, \ and\ \bibinfo {author}
  {\bibfnamefont {J.}~\bibnamefont {Li}},\ }\href {\doibase
  10.1126/science.1256815} {\bibfield  {journal} {\bibinfo  {journal}
  {Science}\ }\textbf {\bibinfo {volume} {346}},\ \bibinfo {pages} {1344}
  (\bibinfo {year} {2014})}\BibitemShut {NoStop}%
\bibitem [{\citenamefont {{Tang}}\ \emph {et~al.}(2017)\citenamefont {{Tang}},
  \citenamefont {{Zhang}}, \citenamefont {{Wong}}, \citenamefont
  {{Pedramrazi}}, \citenamefont {{Tsai}}, \citenamefont {{Jia}}, \citenamefont
  {{Moritz}}, \citenamefont {{Claassen}}, \citenamefont {{Ryu}}, \citenamefont
  {{Kahn}}, \citenamefont {{Jiang}}, \citenamefont {{Yan}}, \citenamefont
  {{Hashimoto}}, \citenamefont {{Lu}}, \citenamefont {{Moore}}, \citenamefont
  {{Hwang}}, \citenamefont {{Hwang}}, \citenamefont {{Hussain}}, \citenamefont
  {{Chen}}, \citenamefont {{Ugeda}}, \citenamefont {{Liu}}, \citenamefont
  {{Xie}}, \citenamefont {{Devereaux}}, \citenamefont {{Crommie}},
  \citenamefont {{Mo}},\ and\ \citenamefont {{Shen}}}]{WTe2QSH_ARPES}%
  \BibitemOpen
  \bibfield  {author} {\bibinfo {author} {\bibfnamefont {S.}~\bibnamefont
  {{Tang}}}, \bibinfo {author} {\bibfnamefont {C.}~\bibnamefont {{Zhang}}},
  \bibinfo {author} {\bibfnamefont {D.}~\bibnamefont {{Wong}}}, \bibinfo
  {author} {\bibfnamefont {Z.}~\bibnamefont {{Pedramrazi}}}, \bibinfo {author}
  {\bibfnamefont {H.-Z.}\ \bibnamefont {{Tsai}}}, \bibinfo {author}
  {\bibfnamefont {C.}~\bibnamefont {{Jia}}}, \bibinfo {author} {\bibfnamefont
  {B.}~\bibnamefont {{Moritz}}}, \bibinfo {author} {\bibfnamefont
  {M.}~\bibnamefont {{Claassen}}}, \bibinfo {author} {\bibfnamefont
  {H.}~\bibnamefont {{Ryu}}}, \bibinfo {author} {\bibfnamefont
  {S.}~\bibnamefont {{Kahn}}}, \bibinfo {author} {\bibfnamefont
  {J.}~\bibnamefont {{Jiang}}}, \bibinfo {author} {\bibfnamefont
  {H.}~\bibnamefont {{Yan}}}, \bibinfo {author} {\bibfnamefont
  {M.}~\bibnamefont {{Hashimoto}}}, \bibinfo {author} {\bibfnamefont
  {D.}~\bibnamefont {{Lu}}}, \bibinfo {author} {\bibfnamefont {R.~G.}\
  \bibnamefont {{Moore}}}, \bibinfo {author} {\bibfnamefont {C.-C.}\
  \bibnamefont {{Hwang}}}, \bibinfo {author} {\bibfnamefont {C.}~\bibnamefont
  {{Hwang}}}, \bibinfo {author} {\bibfnamefont {Z.}~\bibnamefont {{Hussain}}},
  \bibinfo {author} {\bibfnamefont {Y.}~\bibnamefont {{Chen}}}, \bibinfo
  {author} {\bibfnamefont {M.~M.}\ \bibnamefont {{Ugeda}}}, \bibinfo {author}
  {\bibfnamefont {Z.}~\bibnamefont {{Liu}}}, \bibinfo {author} {\bibfnamefont
  {X.}~\bibnamefont {{Xie}}}, \bibinfo {author} {\bibfnamefont {T.~P.}\
  \bibnamefont {{Devereaux}}}, \bibinfo {author} {\bibfnamefont {M.~F.}\
  \bibnamefont {{Crommie}}}, \bibinfo {author} {\bibfnamefont {S.-K.}\
  \bibnamefont {{Mo}}}, \ and\ \bibinfo {author} {\bibfnamefont {Z.-X.}\
  \bibnamefont {{Shen}}},\ }\href {\doibase 10.1038/nphys4174} {\bibfield
  {journal} {\bibinfo  {journal} {Nature Physics}\ }\textbf {\bibinfo {volume}
  {13}},\ \bibinfo {pages} {683} (\bibinfo {year} {2017})}\BibitemShut
  {NoStop}%
\bibitem [{\citenamefont {Fei}\ \emph {et~al.}(2017)\citenamefont {Fei},
  \citenamefont {Palomaki}, \citenamefont {Wu}, \citenamefont {Zhao},
  \citenamefont {Cai}, \citenamefont {Sun}, \citenamefont {Nguyen},
  \citenamefont {Finney}, \citenamefont {Xu},\ and\ \citenamefont
  {Cobden}}]{Exp_WTe2QSH_Cobden}%
  \BibitemOpen
  \bibfield  {author} {\bibinfo {author} {\bibfnamefont {Z.}~\bibnamefont
  {Fei}}, \bibinfo {author} {\bibfnamefont {T.}~\bibnamefont {Palomaki}},
  \bibinfo {author} {\bibfnamefont {S.}~\bibnamefont {Wu}}, \bibinfo {author}
  {\bibfnamefont {W.}~\bibnamefont {Zhao}}, \bibinfo {author} {\bibfnamefont
  {X.}~\bibnamefont {Cai}}, \bibinfo {author} {\bibfnamefont {B.}~\bibnamefont
  {Sun}}, \bibinfo {author} {\bibfnamefont {P.}~\bibnamefont {Nguyen}},
  \bibinfo {author} {\bibfnamefont {J.}~\bibnamefont {Finney}}, \bibinfo
  {author} {\bibfnamefont {X.}~\bibnamefont {Xu}}, \ and\ \bibinfo {author}
  {\bibfnamefont {D.~H.}\ \bibnamefont {Cobden}},\ }\href
  {https://doi.org/10.1038/nphys4091} {\bibfield  {journal} {\bibinfo
  {journal} {Nature Physics}\ }\textbf {\bibinfo {volume} {13}},\ \bibinfo
  {pages} {677} (\bibinfo {year} {2017})}\BibitemShut {NoStop}%
\bibitem [{\citenamefont {{Wu}}\ \emph {et~al.}(2018)\citenamefont {{Wu}},
  \citenamefont {{Fatemi}}, \citenamefont {{Gibson}}, \citenamefont
  {{Watanabe}}, \citenamefont {{Taniguchi}}, \citenamefont {{Cava}},\ and\
  \citenamefont {{Jarillo-Herrero}}}]{Exp_WTe2QSH_Pablo}%
  \BibitemOpen
  \bibfield  {author} {\bibinfo {author} {\bibfnamefont {S.}~\bibnamefont
  {{Wu}}}, \bibinfo {author} {\bibfnamefont {V.}~\bibnamefont {{Fatemi}}},
  \bibinfo {author} {\bibfnamefont {Q.~D.}\ \bibnamefont {{Gibson}}}, \bibinfo
  {author} {\bibfnamefont {K.}~\bibnamefont {{Watanabe}}}, \bibinfo {author}
  {\bibfnamefont {T.}~\bibnamefont {{Taniguchi}}}, \bibinfo {author}
  {\bibfnamefont {R.~J.}\ \bibnamefont {{Cava}}}, \ and\ \bibinfo {author}
  {\bibfnamefont {P.}~\bibnamefont {{Jarillo-Herrero}}},\ }\href {\doibase
  10.1126/science.aan6003} {\bibfield  {journal} {\bibinfo  {journal}
  {Science}\ }\textbf {\bibinfo {volume} {359}},\ \bibinfo {pages} {76}
  (\bibinfo {year} {2018})}\BibitemShut {NoStop}%
\bibitem [{\citenamefont {Li}\ \emph {et~al.}(2017)\citenamefont {Li},
  \citenamefont {Wen}, \citenamefont {He}, \citenamefont {Zhang}, \citenamefont
  {Xia}, \citenamefont {Yu}, \citenamefont {Yang}, \citenamefont {Zhu},
  \citenamefont {Alshareef},\ and\ \citenamefont {Zhang}}]{Exp_WTe2WeylNcomm}%
  \BibitemOpen
  \bibfield  {author} {\bibinfo {author} {\bibfnamefont {P.}~\bibnamefont
  {Li}}, \bibinfo {author} {\bibfnamefont {Y.}~\bibnamefont {Wen}}, \bibinfo
  {author} {\bibfnamefont {X.}~\bibnamefont {He}}, \bibinfo {author}
  {\bibfnamefont {Q.}~\bibnamefont {Zhang}}, \bibinfo {author} {\bibfnamefont
  {C.}~\bibnamefont {Xia}}, \bibinfo {author} {\bibfnamefont {Z.-M.}\
  \bibnamefont {Yu}}, \bibinfo {author} {\bibfnamefont {S.~A.}\ \bibnamefont
  {Yang}}, \bibinfo {author} {\bibfnamefont {Z.}~\bibnamefont {Zhu}}, \bibinfo
  {author} {\bibfnamefont {H.~N.}\ \bibnamefont {Alshareef}}, \ and\ \bibinfo
  {author} {\bibfnamefont {X.-X.}\ \bibnamefont {Zhang}},\ }\href
  {https://doi.org/10.1038/s41467-017-02237-1} {\bibfield  {journal} {\bibinfo
  {journal} {Nature Communications}\ }\textbf {\bibinfo {volume} {8}},\
  \bibinfo {pages} {2150} (\bibinfo {year} {2017})}\BibitemShut {NoStop}%
\bibitem [{\citenamefont {Deng}\ \emph {et~al.}(2016)\citenamefont {Deng},
  \citenamefont {Wan}, \citenamefont {Deng}, \citenamefont {Zhang},
  \citenamefont {Ding}, \citenamefont {Wang}, \citenamefont {Yan},
  \citenamefont {Huang}, \citenamefont {Zhang}, \citenamefont {Xu},
  \citenamefont {Denlinger}, \citenamefont {Fedorov}, \citenamefont {Yang},
  \citenamefont {Duan}, \citenamefont {Yao}, \citenamefont {Wu}, \citenamefont
  {Fan}, \citenamefont {Zhang}, \citenamefont {Chen},\ and\ \citenamefont
  {Zhou}}]{Exp_WeylMoTe2_nphys}%
  \BibitemOpen
  \bibfield  {author} {\bibinfo {author} {\bibfnamefont {K.}~\bibnamefont
  {Deng}}, \bibinfo {author} {\bibfnamefont {G.}~\bibnamefont {Wan}}, \bibinfo
  {author} {\bibfnamefont {P.}~\bibnamefont {Deng}}, \bibinfo {author}
  {\bibfnamefont {K.}~\bibnamefont {Zhang}}, \bibinfo {author} {\bibfnamefont
  {S.}~\bibnamefont {Ding}}, \bibinfo {author} {\bibfnamefont {E.}~\bibnamefont
  {Wang}}, \bibinfo {author} {\bibfnamefont {M.}~\bibnamefont {Yan}}, \bibinfo
  {author} {\bibfnamefont {H.}~\bibnamefont {Huang}}, \bibinfo {author}
  {\bibfnamefont {H.}~\bibnamefont {Zhang}}, \bibinfo {author} {\bibfnamefont
  {Z.}~\bibnamefont {Xu}}, \bibinfo {author} {\bibfnamefont {J.}~\bibnamefont
  {Denlinger}}, \bibinfo {author} {\bibfnamefont {A.}~\bibnamefont {Fedorov}},
  \bibinfo {author} {\bibfnamefont {H.}~\bibnamefont {Yang}}, \bibinfo {author}
  {\bibfnamefont {W.}~\bibnamefont {Duan}}, \bibinfo {author} {\bibfnamefont
  {H.}~\bibnamefont {Yao}}, \bibinfo {author} {\bibfnamefont {Y.}~\bibnamefont
  {Wu}}, \bibinfo {author} {\bibfnamefont {S.}~\bibnamefont {Fan}}, \bibinfo
  {author} {\bibfnamefont {H.}~\bibnamefont {Zhang}}, \bibinfo {author}
  {\bibfnamefont {X.}~\bibnamefont {Chen}}, \ and\ \bibinfo {author}
  {\bibfnamefont {S.}~\bibnamefont {Zhou}},\ }\href
  {https://doi.org/10.1038/nphys3871} {\bibfield  {journal} {\bibinfo
  {journal} {Nature Physics}\ }\textbf {\bibinfo {volume} {12}},\ \bibinfo
  {pages} {1105} (\bibinfo {year} {2016})}\BibitemShut {NoStop}%
\bibitem [{\citenamefont {Castro~Neto}(2001)}]{Thy_CDWTMD_Neto}%
  \BibitemOpen
  \bibfield  {author} {\bibinfo {author} {\bibfnamefont {A.~H.}\ \bibnamefont
  {Castro~Neto}},\ }\href {\doibase 10.1103/PhysRevLett.86.4382} {\bibfield
  {journal} {\bibinfo  {journal} {Phys. Rev. Lett.}\ }\textbf {\bibinfo
  {volume} {86}},\ \bibinfo {pages} {4382} (\bibinfo {year}
  {2001})}\BibitemShut {NoStop}%
\bibitem [{\citenamefont {Wilson}\ \emph {et~al.}(1974)\citenamefont {Wilson},
  \citenamefont {Di~Salvo},\ and\ \citenamefont {Mahajan}}]{Exp_TMDCDW_PRL}%
  \BibitemOpen
  \bibfield  {author} {\bibinfo {author} {\bibfnamefont {J.~A.}\ \bibnamefont
  {Wilson}}, \bibinfo {author} {\bibfnamefont {F.~J.}\ \bibnamefont
  {Di~Salvo}}, \ and\ \bibinfo {author} {\bibfnamefont {S.}~\bibnamefont
  {Mahajan}},\ }\href {\doibase 10.1103/PhysRevLett.32.882} {\bibfield
  {journal} {\bibinfo  {journal} {Phys. Rev. Lett.}\ }\textbf {\bibinfo
  {volume} {32}},\ \bibinfo {pages} {882} (\bibinfo {year} {1974})}\BibitemShut
  {NoStop}%
\bibitem [{\citenamefont {Wilson}\ \emph {et~al.}(2001)\citenamefont {Wilson},
  \citenamefont {Salvo},\ and\ \citenamefont {Mahajan}}]{Exp_NbSe2CDW}%
  \BibitemOpen
  \bibfield  {author} {\bibinfo {author} {\bibfnamefont {J.~A.}\ \bibnamefont
  {Wilson}}, \bibinfo {author} {\bibfnamefont {F.~J.~D.}\ \bibnamefont
  {Salvo}}, \ and\ \bibinfo {author} {\bibfnamefont {S.}~\bibnamefont
  {Mahajan}},\ }\href {\doibase 10.1080/00018730110102718} {\bibfield
  {journal} {\bibinfo  {journal} {Advances in Physics}\ }\textbf {\bibinfo
  {volume} {50}},\ \bibinfo {pages} {1171} (\bibinfo {year} {2001})},\ \Eprint
  {http://arxiv.org/abs/https://doi.org/10.1080/00018730110102718}
  {https://doi.org/10.1080/00018730110102718} \BibitemShut {NoStop}%
\bibitem [{\citenamefont {Sipos}\ \emph {et~al.}(2008)\citenamefont {Sipos},
  \citenamefont {Kusmartseva}, \citenamefont {Akrap}, \citenamefont {Berger},
  \citenamefont {Forró},\ and\ \citenamefont {Tutiš}}]{Exp_TaS2CDW_Nmat}%
  \BibitemOpen
  \bibfield  {author} {\bibinfo {author} {\bibfnamefont {B.}~\bibnamefont
  {Sipos}}, \bibinfo {author} {\bibfnamefont {A.~F.}\ \bibnamefont
  {Kusmartseva}}, \bibinfo {author} {\bibfnamefont {A.}~\bibnamefont {Akrap}},
  \bibinfo {author} {\bibfnamefont {H.}~\bibnamefont {Berger}}, \bibinfo
  {author} {\bibfnamefont {L.}~\bibnamefont {Forró}}, \ and\ \bibinfo {author}
  {\bibfnamefont {E.}~\bibnamefont {Tutiš}},\ }\href
  {https://doi.org/10.1038/nmat2318} {\bibfield  {journal} {\bibinfo  {journal}
  {Nature Materials}\ }\textbf {\bibinfo {volume} {7}},\ \bibinfo {pages} {960}
  (\bibinfo {year} {2008})}\BibitemShut {NoStop}%
\bibitem [{\citenamefont {Xi}\ \emph {et~al.}(2015{\natexlab{a}})\citenamefont
  {Xi}, \citenamefont {Zhao}, \citenamefont {Wang}, \citenamefont {Berger},
  \citenamefont {Forró}, \citenamefont {Shan},\ and\ \citenamefont
  {Mak}}]{Exp_CDWNbSe2_Mak}%
  \BibitemOpen
  \bibfield  {author} {\bibinfo {author} {\bibfnamefont {X.}~\bibnamefont
  {Xi}}, \bibinfo {author} {\bibfnamefont {L.}~\bibnamefont {Zhao}}, \bibinfo
  {author} {\bibfnamefont {Z.}~\bibnamefont {Wang}}, \bibinfo {author}
  {\bibfnamefont {H.}~\bibnamefont {Berger}}, \bibinfo {author} {\bibfnamefont
  {L.}~\bibnamefont {Forró}}, \bibinfo {author} {\bibfnamefont
  {J.}~\bibnamefont {Shan}}, \ and\ \bibinfo {author} {\bibfnamefont {K.~F.}\
  \bibnamefont {Mak}},\ }\href {https://doi.org/10.1038/nnano.2015.143}
  {\bibfield  {journal} {\bibinfo  {journal} {Nature Nanotechnology}\ }\textbf
  {\bibinfo {volume} {10}},\ \bibinfo {pages} {765} (\bibinfo {year}
  {2015}{\natexlab{a}})}\BibitemShut {NoStop}%
\bibitem [{\citenamefont {Ritschel}\ \emph {et~al.}(2015)\citenamefont
  {Ritschel}, \citenamefont {Trinckauf}, \citenamefont {Koepernik},
  \citenamefont {Büchner}, \citenamefont {Zimmermann}, \citenamefont {Berger},
  \citenamefont {Joe}, \citenamefont {Abbamonte},\ and\ \citenamefont
  {Geck}}]{Exp_TaS2CDW_nphys}%
  \BibitemOpen
  \bibfield  {author} {\bibinfo {author} {\bibfnamefont {T.}~\bibnamefont
  {Ritschel}}, \bibinfo {author} {\bibfnamefont {J.}~\bibnamefont {Trinckauf}},
  \bibinfo {author} {\bibfnamefont {K.}~\bibnamefont {Koepernik}}, \bibinfo
  {author} {\bibfnamefont {B.}~\bibnamefont {Büchner}}, \bibinfo {author}
  {\bibfnamefont {M.~v.}\ \bibnamefont {Zimmermann}}, \bibinfo {author}
  {\bibfnamefont {H.}~\bibnamefont {Berger}}, \bibinfo {author} {\bibfnamefont
  {Y.~I.}\ \bibnamefont {Joe}}, \bibinfo {author} {\bibfnamefont
  {P.}~\bibnamefont {Abbamonte}}, \ and\ \bibinfo {author} {\bibfnamefont
  {J.}~\bibnamefont {Geck}},\ }\href {https://doi.org/10.1038/nphys3267}
  {\bibfield  {journal} {\bibinfo  {journal} {Nature Physics}\ }\textbf
  {\bibinfo {volume} {11}},\ \bibinfo {pages} {328} (\bibinfo {year}
  {2015})}\BibitemShut {NoStop}%
\bibitem [{\citenamefont {Li}\ \emph {et~al.}(2015)\citenamefont {Li},
  \citenamefont {O’Farrell}, \citenamefont {Loh}, \citenamefont {Eda},
  \citenamefont {Özyilmaz},\ and\ \citenamefont
  {Castro~Neto}}]{Exp_GateTunedTiSe2_Neto}%
  \BibitemOpen
  \bibfield  {author} {\bibinfo {author} {\bibfnamefont {L.~J.}\ \bibnamefont
  {Li}}, \bibinfo {author} {\bibfnamefont {E.~C.~T.}\ \bibnamefont
  {O’Farrell}}, \bibinfo {author} {\bibfnamefont {K.~P.}\ \bibnamefont
  {Loh}}, \bibinfo {author} {\bibfnamefont {G.}~\bibnamefont {Eda}}, \bibinfo
  {author} {\bibfnamefont {B.}~\bibnamefont {Özyilmaz}}, \ and\ \bibinfo
  {author} {\bibfnamefont {A.~H.}\ \bibnamefont {Castro~Neto}},\ }\href
  {https://doi.org/10.1038/nature16175} {\bibfield  {journal} {\bibinfo
  {journal} {Nature}\ }\textbf {\bibinfo {volume} {529}},\ \bibinfo {pages}
  {185} (\bibinfo {year} {2015})}\BibitemShut {NoStop}%
\bibitem [{\citenamefont {Yuan}\ \emph {et~al.}(2014)\citenamefont {Yuan},
  \citenamefont {Mak},\ and\ \citenamefont {Law}}]{Toposc_MoS2_Law}%
  \BibitemOpen
  \bibfield  {author} {\bibinfo {author} {\bibfnamefont {N.~F.~Q.}\
  \bibnamefont {Yuan}}, \bibinfo {author} {\bibfnamefont {K.~F.}\ \bibnamefont
  {Mak}}, \ and\ \bibinfo {author} {\bibfnamefont {K.~T.}\ \bibnamefont
  {Law}},\ }\href {\doibase 10.1103/PhysRevLett.113.097001} {\bibfield
  {journal} {\bibinfo  {journal} {Phys. Rev. Lett.}\ }\textbf {\bibinfo
  {volume} {113}},\ \bibinfo {pages} {097001} (\bibinfo {year}
  {2014})}\BibitemShut {NoStop}%
\bibitem [{\citenamefont {Hsu}\ \emph {et~al.}(2017)\citenamefont {Hsu},
  \citenamefont {Vaezi}, \citenamefont {Fischer},\ and\ \citenamefont
  {Kim}}]{Toposc_holeMoS2_Hsu}%
  \BibitemOpen
  \bibfield  {author} {\bibinfo {author} {\bibfnamefont {Y.-T.}\ \bibnamefont
  {Hsu}}, \bibinfo {author} {\bibfnamefont {A.}~\bibnamefont {Vaezi}}, \bibinfo
  {author} {\bibfnamefont {M.~H.}\ \bibnamefont {Fischer}}, \ and\ \bibinfo
  {author} {\bibfnamefont {E.-A.}\ \bibnamefont {Kim}},\ }\href
  {https://doi.org/10.1038/ncomms14985} {\bibfield  {journal} {\bibinfo
  {journal} {Nature Communications}\ }\textbf {\bibinfo {volume} {8}},\
  \bibinfo {pages} {14985} (\bibinfo {year} {2017})}\BibitemShut {NoStop}%
\bibitem [{\citenamefont {Ye}\ \emph {et~al.}(2012)\citenamefont {Ye},
  \citenamefont {Zhang}, \citenamefont {Akashi}, \citenamefont {Bahramy},
  \citenamefont {Arita},\ and\ \citenamefont {Iwasa}}]{Exp_MoS2sc_Iwasa}%
  \BibitemOpen
  \bibfield  {author} {\bibinfo {author} {\bibfnamefont {J.~T.}\ \bibnamefont
  {Ye}}, \bibinfo {author} {\bibfnamefont {Y.~J.}\ \bibnamefont {Zhang}},
  \bibinfo {author} {\bibfnamefont {R.}~\bibnamefont {Akashi}}, \bibinfo
  {author} {\bibfnamefont {M.~S.}\ \bibnamefont {Bahramy}}, \bibinfo {author}
  {\bibfnamefont {R.}~\bibnamefont {Arita}}, \ and\ \bibinfo {author}
  {\bibfnamefont {Y.}~\bibnamefont {Iwasa}},\ }\href {\doibase
  10.1126/science.1228006} {\bibfield  {journal} {\bibinfo  {journal}
  {Science}\ }\textbf {\bibinfo {volume} {338}},\ \bibinfo {pages} {1193}
  (\bibinfo {year} {2012})},\ \Eprint
  {http://arxiv.org/abs/http://science.sciencemag.org/content/338/6111/1193.full.pdf}
  {http://science.sciencemag.org/content/338/6111/1193.full.pdf} \BibitemShut
  {NoStop}%
\bibitem [{\citenamefont {Lu}\ \emph {et~al.}(2015)\citenamefont {Lu},
  \citenamefont {Zheliuk}, \citenamefont {Leermakers}, \citenamefont {Yuan},
  \citenamefont {Zeitler}, \citenamefont {Law},\ and\ \citenamefont
  {Ye}}]{Exp_IsingSc_Ye}%
  \BibitemOpen
  \bibfield  {author} {\bibinfo {author} {\bibfnamefont {J.~M.}\ \bibnamefont
  {Lu}}, \bibinfo {author} {\bibfnamefont {O.}~\bibnamefont {Zheliuk}},
  \bibinfo {author} {\bibfnamefont {I.}~\bibnamefont {Leermakers}}, \bibinfo
  {author} {\bibfnamefont {N.~F.~Q.}\ \bibnamefont {Yuan}}, \bibinfo {author}
  {\bibfnamefont {U.}~\bibnamefont {Zeitler}}, \bibinfo {author} {\bibfnamefont
  {K.~T.}\ \bibnamefont {Law}}, \ and\ \bibinfo {author} {\bibfnamefont
  {J.~T.}\ \bibnamefont {Ye}},\ }\href {\doibase 10.1126/science.aab2277}
  {\bibfield  {journal} {\bibinfo  {journal} {Science}\ }\textbf {\bibinfo
  {volume} {350}},\ \bibinfo {pages} {1353} (\bibinfo {year} {2015})},\ \Eprint
  {http://arxiv.org/abs/http://science.sciencemag.org/content/350/6266/1353.full.pdf}
  {http://science.sciencemag.org/content/350/6266/1353.full.pdf} \BibitemShut
  {NoStop}%
\bibitem [{\citenamefont {Shi}\ \emph {et~al.}(2015)\citenamefont {Shi},
  \citenamefont {Ye}, \citenamefont {Zhang}, \citenamefont {Suzuki},
  \citenamefont {Yoshida}, \citenamefont {Miyazaki}, \citenamefont {Inoue},
  \citenamefont {Saito},\ and\ \citenamefont {Iwasa}}]{Exp_TMDsc_Iwasa}%
  \BibitemOpen
  \bibfield  {author} {\bibinfo {author} {\bibfnamefont {W.}~\bibnamefont
  {Shi}}, \bibinfo {author} {\bibfnamefont {J.}~\bibnamefont {Ye}}, \bibinfo
  {author} {\bibfnamefont {Y.}~\bibnamefont {Zhang}}, \bibinfo {author}
  {\bibfnamefont {R.}~\bibnamefont {Suzuki}}, \bibinfo {author} {\bibfnamefont
  {M.}~\bibnamefont {Yoshida}}, \bibinfo {author} {\bibfnamefont
  {J.}~\bibnamefont {Miyazaki}}, \bibinfo {author} {\bibfnamefont
  {N.}~\bibnamefont {Inoue}}, \bibinfo {author} {\bibfnamefont
  {Y.}~\bibnamefont {Saito}}, \ and\ \bibinfo {author} {\bibfnamefont
  {Y.}~\bibnamefont {Iwasa}},\ }\href {https://doi.org/10.1038/srep12534}
  {\bibfield  {journal} {\bibinfo  {journal} {Scientific Reports}\ }\textbf
  {\bibinfo {volume} {5}},\ \bibinfo {pages} {12534} (\bibinfo {year}
  {2015})}\BibitemShut {NoStop}%
\bibitem [{\citenamefont {Xi}\ \emph {et~al.}(2015{\natexlab{b}})\citenamefont
  {Xi}, \citenamefont {Wang}, \citenamefont {Zhao}, \citenamefont {Park},
  \citenamefont {Law}, \citenamefont {Berger}, \citenamefont {Forró},
  \citenamefont {Shan},\ and\ \citenamefont {Mak}}]{Exp_NbSe2sc_Mak}%
  \BibitemOpen
  \bibfield  {author} {\bibinfo {author} {\bibfnamefont {X.}~\bibnamefont
  {Xi}}, \bibinfo {author} {\bibfnamefont {Z.}~\bibnamefont {Wang}}, \bibinfo
  {author} {\bibfnamefont {W.}~\bibnamefont {Zhao}}, \bibinfo {author}
  {\bibfnamefont {J.-H.}\ \bibnamefont {Park}}, \bibinfo {author}
  {\bibfnamefont {K.~T.}\ \bibnamefont {Law}}, \bibinfo {author} {\bibfnamefont
  {H.}~\bibnamefont {Berger}}, \bibinfo {author} {\bibfnamefont
  {L.}~\bibnamefont {Forró}}, \bibinfo {author} {\bibfnamefont
  {J.}~\bibnamefont {Shan}}, \ and\ \bibinfo {author} {\bibfnamefont {K.~F.}\
  \bibnamefont {Mak}},\ }\href {https://doi.org/10.1038/nphys3538} {\bibfield
  {journal} {\bibinfo  {journal} {Nature Physics}\ }\textbf {\bibinfo {volume}
  {12}},\ \bibinfo {pages} {139} (\bibinfo {year}
  {2015}{\natexlab{b}})}\BibitemShut {NoStop}%
\bibitem [{\citenamefont {Sajadi}\ \emph {et~al.}(2018)\citenamefont {Sajadi},
  \citenamefont {Palomaki}, \citenamefont {Fei}, \citenamefont {Zhao},
  \citenamefont {Bement}, \citenamefont {Olsen}, \citenamefont {Luescher},
  \citenamefont {Xu}, \citenamefont {Folk},\ and\ \citenamefont
  {Cobden}}]{Exp_WTe2sc_Cobden}%
  \BibitemOpen
  \bibfield  {author} {\bibinfo {author} {\bibfnamefont {E.}~\bibnamefont
  {Sajadi}}, \bibinfo {author} {\bibfnamefont {T.}~\bibnamefont {Palomaki}},
  \bibinfo {author} {\bibfnamefont {Z.}~\bibnamefont {Fei}}, \bibinfo {author}
  {\bibfnamefont {W.}~\bibnamefont {Zhao}}, \bibinfo {author} {\bibfnamefont
  {P.}~\bibnamefont {Bement}}, \bibinfo {author} {\bibfnamefont
  {C.}~\bibnamefont {Olsen}}, \bibinfo {author} {\bibfnamefont
  {S.}~\bibnamefont {Luescher}}, \bibinfo {author} {\bibfnamefont
  {X.}~\bibnamefont {Xu}}, \bibinfo {author} {\bibfnamefont {J.~A.}\
  \bibnamefont {Folk}}, \ and\ \bibinfo {author} {\bibfnamefont {D.~H.}\
  \bibnamefont {Cobden}},\ }\href {\doibase 10.1126/science.aar4426} {\bibfield
   {journal} {\bibinfo  {journal} {Science}\ }\textbf {\bibinfo {volume}
  {362}},\ \bibinfo {pages} {922} (\bibinfo {year} {2018})},\ \Eprint
  {http://arxiv.org/abs/http://science.sciencemag.org/content/362/6417/922.full.pdf}
  {http://science.sciencemag.org/content/362/6417/922.full.pdf} \BibitemShut
  {NoStop}%
\bibitem [{\citenamefont {Fatemi}\ \emph {et~al.}(2018)\citenamefont {Fatemi},
  \citenamefont {Wu}, \citenamefont {Cao}, \citenamefont {Bretheau},
  \citenamefont {Gibson}, \citenamefont {Watanabe}, \citenamefont {Taniguchi},
  \citenamefont {Cava},\ and\ \citenamefont
  {Jarillo-Herrero}}]{Exp_WTe2sc_Pablo}%
  \BibitemOpen
  \bibfield  {author} {\bibinfo {author} {\bibfnamefont {V.}~\bibnamefont
  {Fatemi}}, \bibinfo {author} {\bibfnamefont {S.}~\bibnamefont {Wu}}, \bibinfo
  {author} {\bibfnamefont {Y.}~\bibnamefont {Cao}}, \bibinfo {author}
  {\bibfnamefont {L.}~\bibnamefont {Bretheau}}, \bibinfo {author}
  {\bibfnamefont {Q.~D.}\ \bibnamefont {Gibson}}, \bibinfo {author}
  {\bibfnamefont {K.}~\bibnamefont {Watanabe}}, \bibinfo {author}
  {\bibfnamefont {T.}~\bibnamefont {Taniguchi}}, \bibinfo {author}
  {\bibfnamefont {R.~J.}\ \bibnamefont {Cava}}, \ and\ \bibinfo {author}
  {\bibfnamefont {P.}~\bibnamefont {Jarillo-Herrero}},\ }\href {\doibase
  10.1126/science.aar4642} {\bibfield  {journal} {\bibinfo  {journal}
  {Science}\ }\textbf {\bibinfo {volume} {362}},\ \bibinfo {pages} {926}
  (\bibinfo {year} {2018})},\ \Eprint
  {http://arxiv.org/abs/http://science.sciencemag.org/content/362/6417/926.full.pdf}
  {http://science.sciencemag.org/content/362/6417/926.full.pdf} \BibitemShut
  {NoStop}%
\bibitem [{\citenamefont {Kusmartseva}\ \emph {et~al.}(2009)\citenamefont
  {Kusmartseva}, \citenamefont {Sipos}, \citenamefont {Berger}, \citenamefont
  {Forr\'o},\ and\ \citenamefont {Tuti\ifmmode~\check{s}\else
  \v{s}\fi{}}}]{Exp_TiSe2sc_pressure}%
  \BibitemOpen
  \bibfield  {author} {\bibinfo {author} {\bibfnamefont {A.~F.}\ \bibnamefont
  {Kusmartseva}}, \bibinfo {author} {\bibfnamefont {B.}~\bibnamefont {Sipos}},
  \bibinfo {author} {\bibfnamefont {H.}~\bibnamefont {Berger}}, \bibinfo
  {author} {\bibfnamefont {L.}~\bibnamefont {Forr\'o}}, \ and\ \bibinfo
  {author} {\bibfnamefont {E.}~\bibnamefont {Tuti\ifmmode~\check{s}\else
  \v{s}\fi{}}},\ }\href {\doibase 10.1103/PhysRevLett.103.236401} {\bibfield
  {journal} {\bibinfo  {journal} {Phys. Rev. Lett.}\ }\textbf {\bibinfo
  {volume} {103}},\ \bibinfo {pages} {236401} (\bibinfo {year}
  {2009})}\BibitemShut {NoStop}%
\bibitem [{\citenamefont {Kang}\ \emph {et~al.}(2015)\citenamefont {Kang},
  \citenamefont {Zhou}, \citenamefont {Yi}, \citenamefont {Yang}, \citenamefont
  {Guo}, \citenamefont {Shi}, \citenamefont {Zhang}, \citenamefont {Wang},
  \citenamefont {Zhang}, \citenamefont {Jiang}, \citenamefont {Li},
  \citenamefont {Yang}, \citenamefont {Wu}, \citenamefont {Zhang},
  \citenamefont {Sun},\ and\ \citenamefont {Zhao}}]{Exp_WTe2_pressure}%
  \BibitemOpen
  \bibfield  {author} {\bibinfo {author} {\bibfnamefont {D.}~\bibnamefont
  {Kang}}, \bibinfo {author} {\bibfnamefont {Y.}~\bibnamefont {Zhou}}, \bibinfo
  {author} {\bibfnamefont {W.}~\bibnamefont {Yi}}, \bibinfo {author}
  {\bibfnamefont {C.}~\bibnamefont {Yang}}, \bibinfo {author} {\bibfnamefont
  {J.}~\bibnamefont {Guo}}, \bibinfo {author} {\bibfnamefont {Y.}~\bibnamefont
  {Shi}}, \bibinfo {author} {\bibfnamefont {S.}~\bibnamefont {Zhang}}, \bibinfo
  {author} {\bibfnamefont {Z.}~\bibnamefont {Wang}}, \bibinfo {author}
  {\bibfnamefont {C.}~\bibnamefont {Zhang}}, \bibinfo {author} {\bibfnamefont
  {S.}~\bibnamefont {Jiang}}, \bibinfo {author} {\bibfnamefont
  {A.}~\bibnamefont {Li}}, \bibinfo {author} {\bibfnamefont {K.}~\bibnamefont
  {Yang}}, \bibinfo {author} {\bibfnamefont {Q.}~\bibnamefont {Wu}}, \bibinfo
  {author} {\bibfnamefont {G.}~\bibnamefont {Zhang}}, \bibinfo {author}
  {\bibfnamefont {L.}~\bibnamefont {Sun}}, \ and\ \bibinfo {author}
  {\bibfnamefont {Z.}~\bibnamefont {Zhao}},\ }\href
  {https://doi.org/10.1038/ncomms8804} {\bibfield  {journal} {\bibinfo
  {journal} {Nature Communications}\ }\textbf {\bibinfo {volume} {6}},\
  \bibinfo {pages} {7804} (\bibinfo {year} {2015})}\BibitemShut {NoStop}%
\bibitem [{\citenamefont {Pan}\ \emph {et~al.}(2015)\citenamefont {Pan},
  \citenamefont {Chen}, \citenamefont {Liu}, \citenamefont {Feng},
  \citenamefont {Wei}, \citenamefont {Zhou}, \citenamefont {Chi}, \citenamefont
  {Pi}, \citenamefont {Yen}, \citenamefont {Song}, \citenamefont {Wan},
  \citenamefont {Yang}, \citenamefont {Wang}, \citenamefont {Wang},\ and\
  \citenamefont {Zhang}}]{Exp_WTe2Pressure_Pan}%
  \BibitemOpen
  \bibfield  {author} {\bibinfo {author} {\bibfnamefont {X.-C.}\ \bibnamefont
  {Pan}}, \bibinfo {author} {\bibfnamefont {X.}~\bibnamefont {Chen}}, \bibinfo
  {author} {\bibfnamefont {H.}~\bibnamefont {Liu}}, \bibinfo {author}
  {\bibfnamefont {Y.}~\bibnamefont {Feng}}, \bibinfo {author} {\bibfnamefont
  {Z.}~\bibnamefont {Wei}}, \bibinfo {author} {\bibfnamefont {Y.}~\bibnamefont
  {Zhou}}, \bibinfo {author} {\bibfnamefont {Z.}~\bibnamefont {Chi}}, \bibinfo
  {author} {\bibfnamefont {L.}~\bibnamefont {Pi}}, \bibinfo {author}
  {\bibfnamefont {F.}~\bibnamefont {Yen}}, \bibinfo {author} {\bibfnamefont
  {F.}~\bibnamefont {Song}}, \bibinfo {author} {\bibfnamefont {X.}~\bibnamefont
  {Wan}}, \bibinfo {author} {\bibfnamefont {Z.}~\bibnamefont {Yang}}, \bibinfo
  {author} {\bibfnamefont {B.}~\bibnamefont {Wang}}, \bibinfo {author}
  {\bibfnamefont {G.}~\bibnamefont {Wang}}, \ and\ \bibinfo {author}
  {\bibfnamefont {Y.}~\bibnamefont {Zhang}},\ }\href
  {https://doi.org/10.1038/ncomms8805} {\bibfield  {journal} {\bibinfo
  {journal} {Nature Communications}\ }\textbf {\bibinfo {volume} {6}},\
  \bibinfo {pages} {7805} (\bibinfo {year} {2015})}\BibitemShut {NoStop}%
\bibitem [{\citenamefont {Chi}\ \emph {et~al.}(2018)\citenamefont {Chi},
  \citenamefont {Chen}, \citenamefont {Yen}, \citenamefont {Peng},
  \citenamefont {Zhou}, \citenamefont {Zhu}, \citenamefont {Zhang},
  \citenamefont {Liu}, \citenamefont {Lin}, \citenamefont {Chu}, \citenamefont
  {Li}, \citenamefont {Zhao}, \citenamefont {Kagayama}, \citenamefont {Ma},\
  and\ \citenamefont {Yang}}]{Exp_MoS2sc_Pressure}%
  \BibitemOpen
  \bibfield  {author} {\bibinfo {author} {\bibfnamefont {Z.}~\bibnamefont
  {Chi}}, \bibinfo {author} {\bibfnamefont {X.}~\bibnamefont {Chen}}, \bibinfo
  {author} {\bibfnamefont {F.}~\bibnamefont {Yen}}, \bibinfo {author}
  {\bibfnamefont {F.}~\bibnamefont {Peng}}, \bibinfo {author} {\bibfnamefont
  {Y.}~\bibnamefont {Zhou}}, \bibinfo {author} {\bibfnamefont {J.}~\bibnamefont
  {Zhu}}, \bibinfo {author} {\bibfnamefont {Y.}~\bibnamefont {Zhang}}, \bibinfo
  {author} {\bibfnamefont {X.}~\bibnamefont {Liu}}, \bibinfo {author}
  {\bibfnamefont {C.}~\bibnamefont {Lin}}, \bibinfo {author} {\bibfnamefont
  {S.}~\bibnamefont {Chu}}, \bibinfo {author} {\bibfnamefont {Y.}~\bibnamefont
  {Li}}, \bibinfo {author} {\bibfnamefont {J.}~\bibnamefont {Zhao}}, \bibinfo
  {author} {\bibfnamefont {T.}~\bibnamefont {Kagayama}}, \bibinfo {author}
  {\bibfnamefont {Y.}~\bibnamefont {Ma}}, \ and\ \bibinfo {author}
  {\bibfnamefont {Z.}~\bibnamefont {Yang}},\ }\href {\doibase
  10.1103/PhysRevLett.120.037002} {\bibfield  {journal} {\bibinfo  {journal}
  {Phys. Rev. Lett.}\ }\textbf {\bibinfo {volume} {120}},\ \bibinfo {pages}
  {037002} (\bibinfo {year} {2018})}\BibitemShut {NoStop}%
\bibitem [{\citenamefont {Yu}\ \emph {et~al.}(2015)\citenamefont {Yu},
  \citenamefont {Yang}, \citenamefont {Lu}, \citenamefont {Yan}, \citenamefont
  {Cho}, \citenamefont {Ma}, \citenamefont {Niu}, \citenamefont {Kim},
  \citenamefont {Son}, \citenamefont {Feng}, \citenamefont {Li}, \citenamefont
  {Cheong}, \citenamefont {Chen},\ and\ \citenamefont {Zhang}}]{Exp_TaS2gate}%
  \BibitemOpen
  \bibfield  {author} {\bibinfo {author} {\bibfnamefont {Y.}~\bibnamefont
  {Yu}}, \bibinfo {author} {\bibfnamefont {F.}~\bibnamefont {Yang}}, \bibinfo
  {author} {\bibfnamefont {X.~F.}\ \bibnamefont {Lu}}, \bibinfo {author}
  {\bibfnamefont {Y.~J.}\ \bibnamefont {Yan}}, \bibinfo {author} {\bibfnamefont
  {Y.-H.}\ \bibnamefont {Cho}}, \bibinfo {author} {\bibfnamefont
  {L.}~\bibnamefont {Ma}}, \bibinfo {author} {\bibfnamefont {X.}~\bibnamefont
  {Niu}}, \bibinfo {author} {\bibfnamefont {S.}~\bibnamefont {Kim}}, \bibinfo
  {author} {\bibfnamefont {Y.-W.}\ \bibnamefont {Son}}, \bibinfo {author}
  {\bibfnamefont {D.}~\bibnamefont {Feng}}, \bibinfo {author} {\bibfnamefont
  {S.}~\bibnamefont {Li}}, \bibinfo {author} {\bibfnamefont {S.-W.}\
  \bibnamefont {Cheong}}, \bibinfo {author} {\bibfnamefont {X.~H.}\
  \bibnamefont {Chen}}, \ and\ \bibinfo {author} {\bibfnamefont
  {Y.}~\bibnamefont {Zhang}},\ }\href {https://doi.org/10.1038/nnano.2014.323}
  {\bibfield  {journal} {\bibinfo  {journal} {Nature Nanotechnology}\ }\textbf
  {\bibinfo {volume} {10}},\ \bibinfo {pages} {270} (\bibinfo {year}
  {2015})}\BibitemShut {NoStop}%
\bibitem [{\citenamefont {Wu}\ \emph {et~al.}(2018)\citenamefont {Wu},
  \citenamefont {Lovorn}, \citenamefont {Tutuc},\ and\ \citenamefont
  {MacDonald}}]{WSe2_Moire_Wu}%
  \BibitemOpen
  \bibfield  {author} {\bibinfo {author} {\bibfnamefont {F.}~\bibnamefont
  {Wu}}, \bibinfo {author} {\bibfnamefont {T.}~\bibnamefont {Lovorn}}, \bibinfo
  {author} {\bibfnamefont {E.}~\bibnamefont {Tutuc}}, \ and\ \bibinfo {author}
  {\bibfnamefont {A.~H.}\ \bibnamefont {MacDonald}},\ }\href {\doibase
  10.1103/PhysRevLett.121.026402} {\bibfield  {journal} {\bibinfo  {journal}
  {Phys. Rev. Lett.}\ }\textbf {\bibinfo {volume} {121}},\ \bibinfo {pages}
  {026402} (\bibinfo {year} {2018})}\BibitemShut {NoStop}%
\bibitem [{\citenamefont {{Wu}}\ \emph {et~al.}(2018)\citenamefont {{Wu}},
  \citenamefont {{Lovorn}}, \citenamefont {{Tutuc}}, \citenamefont {{Martin}},\
  and\ \citenamefont {{MacDonald}}}]{MoireHomo_MoTe2_Wu}%
  \BibitemOpen
  \bibfield  {author} {\bibinfo {author} {\bibfnamefont {F.}~\bibnamefont
  {{Wu}}}, \bibinfo {author} {\bibfnamefont {T.}~\bibnamefont {{Lovorn}}},
  \bibinfo {author} {\bibfnamefont {E.}~\bibnamefont {{Tutuc}}}, \bibinfo
  {author} {\bibfnamefont {I.}~\bibnamefont {{Martin}}}, \ and\ \bibinfo
  {author} {\bibfnamefont {A.~H.}\ \bibnamefont {{MacDonald}}},\ }\href@noop {}
  {\bibfield  {journal} {\bibinfo  {journal} {arXiv e-prints}\ ,\ \bibinfo
  {eid} {arXiv:1807.03311}} (\bibinfo {year} {2018})},\ \Eprint
  {http://arxiv.org/abs/1807.03311} {arXiv:1807.03311 [cond-mat.mes-hall]}
  \BibitemShut {NoStop}%
\bibitem [{\citenamefont {{Jia}}\ \emph {et~al.}(2017)\citenamefont {{Jia}},
  \citenamefont {{Song}}, \citenamefont {{Li}}, \citenamefont {{Ran}},
  \citenamefont {{Lu}}, \citenamefont {{Zheng}}, \citenamefont {{Zhu}},
  \citenamefont {{Shi}}, \citenamefont {{Sun}}, \citenamefont {{Wen}},
  \citenamefont {{Xing}},\ and\ \citenamefont {{Li}}}]{jia_2017prb}%
  \BibitemOpen
  \bibfield  {author} {\bibinfo {author} {\bibfnamefont {Z.-Y.}\ \bibnamefont
  {{Jia}}}, \bibinfo {author} {\bibfnamefont {Y.-H.}\ \bibnamefont {{Song}}},
  \bibinfo {author} {\bibfnamefont {X.-B.}\ \bibnamefont {{Li}}}, \bibinfo
  {author} {\bibfnamefont {K.}~\bibnamefont {{Ran}}}, \bibinfo {author}
  {\bibfnamefont {P.}~\bibnamefont {{Lu}}}, \bibinfo {author} {\bibfnamefont
  {H.-J.}\ \bibnamefont {{Zheng}}}, \bibinfo {author} {\bibfnamefont {X.-Y.}\
  \bibnamefont {{Zhu}}}, \bibinfo {author} {\bibfnamefont {Z.-Q.}\ \bibnamefont
  {{Shi}}}, \bibinfo {author} {\bibfnamefont {J.}~\bibnamefont {{Sun}}},
  \bibinfo {author} {\bibfnamefont {J.}~\bibnamefont {{Wen}}}, \bibinfo
  {author} {\bibfnamefont {D.}~\bibnamefont {{Xing}}}, \ and\ \bibinfo {author}
  {\bibfnamefont {S.-C.}\ \bibnamefont {{Li}}},\ }\href {\doibase
  10.1103/PhysRevB.96.041108} {\bibfield  {journal} {\bibinfo  {journal}
  {\prb}\ }\textbf {\bibinfo {volume} {96}},\ \bibinfo {eid} {041108} (\bibinfo
  {year} {2017})}\BibitemShut {NoStop}%
\bibitem [{\citenamefont {{Peng}}\ \emph {et~al.}(2017)\citenamefont {{Peng}},
  \citenamefont {{Yuan}}, \citenamefont {{Li}}, \citenamefont {{Yang}},
  \citenamefont {{Xian}}, \citenamefont {{Yi}}, \citenamefont {{Shi}},\ and\
  \citenamefont {{Fu}}}]{peng_2017ncomm}%
  \BibitemOpen
  \bibfield  {author} {\bibinfo {author} {\bibfnamefont {L.}~\bibnamefont
  {{Peng}}}, \bibinfo {author} {\bibfnamefont {Y.}~\bibnamefont {{Yuan}}},
  \bibinfo {author} {\bibfnamefont {G.}~\bibnamefont {{Li}}}, \bibinfo {author}
  {\bibfnamefont {X.}~\bibnamefont {{Yang}}}, \bibinfo {author} {\bibfnamefont
  {J.-J.}\ \bibnamefont {{Xian}}}, \bibinfo {author} {\bibfnamefont {C.-J.}\
  \bibnamefont {{Yi}}}, \bibinfo {author} {\bibfnamefont {Y.-G.}\ \bibnamefont
  {{Shi}}}, \ and\ \bibinfo {author} {\bibfnamefont {Y.-S.}\ \bibnamefont
  {{Fu}}},\ }\href {\doibase 10.1038/s41467-017-00745-8} {\bibfield  {journal}
  {\bibinfo  {journal} {Nature Communications}\ }\textbf {\bibinfo {volume}
  {8}},\ \bibinfo {eid} {659} (\bibinfo {year} {2017})}\BibitemShut {NoStop}%
\bibitem [{\citenamefont {Kane}\ and\ \citenamefont
  {Mele}(2005{\natexlab{a}})}]{QSH_KaneMele}%
  \BibitemOpen
  \bibfield  {author} {\bibinfo {author} {\bibfnamefont {C.~L.}\ \bibnamefont
  {Kane}}\ and\ \bibinfo {author} {\bibfnamefont {E.~J.}\ \bibnamefont
  {Mele}},\ }\href {\doibase 10.1103/PhysRevLett.95.226801} {\bibfield
  {journal} {\bibinfo  {journal} {Phys. Rev. Lett.}\ }\textbf {\bibinfo
  {volume} {95}},\ \bibinfo {pages} {226801} (\bibinfo {year}
  {2005}{\natexlab{a}})}\BibitemShut {NoStop}%
\bibitem [{\citenamefont {Bernevig}\ and\ \citenamefont
  {Zhang}(2006)}]{QSH_Bernevig}%
  \BibitemOpen
  \bibfield  {author} {\bibinfo {author} {\bibfnamefont {B.~A.}\ \bibnamefont
  {Bernevig}}\ and\ \bibinfo {author} {\bibfnamefont {S.-C.}\ \bibnamefont
  {Zhang}},\ }\href {\doibase 10.1103/PhysRevLett.96.106802} {\bibfield
  {journal} {\bibinfo  {journal} {Phys. Rev. Lett.}\ }\textbf {\bibinfo
  {volume} {96}},\ \bibinfo {pages} {106802} (\bibinfo {year}
  {2006})}\BibitemShut {NoStop}%
\bibitem [{\citenamefont {Qi}\ \emph {et~al.}(2010)\citenamefont {Qi},
  \citenamefont {Hughes},\ and\ \citenamefont {Zhang}}]{IndexTRsc_Zhang}%
  \BibitemOpen
  \bibfield  {author} {\bibinfo {author} {\bibfnamefont {X.-L.}\ \bibnamefont
  {Qi}}, \bibinfo {author} {\bibfnamefont {T.~L.}\ \bibnamefont {Hughes}}, \
  and\ \bibinfo {author} {\bibfnamefont {S.-C.}\ \bibnamefont {Zhang}},\ }\href
  {\doibase 10.1103/PhysRevB.81.134508} {\bibfield  {journal} {\bibinfo
  {journal} {Phys. Rev. B}\ }\textbf {\bibinfo {volume} {81}},\ \bibinfo
  {pages} {134508} (\bibinfo {year} {2010})}\BibitemShut {NoStop}%
\bibitem [{\citenamefont {Zhang}\ \emph {et~al.}(2013)\citenamefont {Zhang},
  \citenamefont {Kane},\ and\ \citenamefont {Mele}}]{TscMajo_FanZhang}%
  \BibitemOpen
  \bibfield  {author} {\bibinfo {author} {\bibfnamefont {F.}~\bibnamefont
  {Zhang}}, \bibinfo {author} {\bibfnamefont {C.~L.}\ \bibnamefont {Kane}}, \
  and\ \bibinfo {author} {\bibfnamefont {E.~J.}\ \bibnamefont {Mele}},\ }\href
  {\doibase 10.1103/PhysRevLett.111.056402} {\bibfield  {journal} {\bibinfo
  {journal} {Phys. Rev. Lett.}\ }\textbf {\bibinfo {volume} {111}},\ \bibinfo
  {pages} {056402} (\bibinfo {year} {2013})}\BibitemShut {NoStop}%
\bibitem [{\citenamefont {Khalaf}\ \emph {et~al.}(2018)\citenamefont {Khalaf},
  \citenamefont {Po}, \citenamefont {Vishwanath},\ and\ \citenamefont
  {Watanabe}}]{Indicator_PRX}%
  \BibitemOpen
  \bibfield  {author} {\bibinfo {author} {\bibfnamefont {E.}~\bibnamefont
  {Khalaf}}, \bibinfo {author} {\bibfnamefont {H.~C.}\ \bibnamefont {Po}},
  \bibinfo {author} {\bibfnamefont {A.}~\bibnamefont {Vishwanath}}, \ and\
  \bibinfo {author} {\bibfnamefont {H.}~\bibnamefont {Watanabe}},\ }\href
  {\doibase 10.1103/PhysRevX.8.031070} {\bibfield  {journal} {\bibinfo
  {journal} {Phys. Rev. X}\ }\textbf {\bibinfo {volume} {8}},\ \bibinfo {pages}
  {031070} (\bibinfo {year} {2018})}\BibitemShut {NoStop}%
\bibitem [{\citenamefont {Khalaf}(2018)}]{InvHOTsc_Khalaf}%
  \BibitemOpen
  \bibfield  {author} {\bibinfo {author} {\bibfnamefont {E.}~\bibnamefont
  {Khalaf}},\ }\href {\doibase 10.1103/PhysRevB.97.205136} {\bibfield
  {journal} {\bibinfo  {journal} {Phys. Rev. B}\ }\textbf {\bibinfo {volume}
  {97}},\ \bibinfo {pages} {205136} (\bibinfo {year} {2018})}\BibitemShut
  {NoStop}%
\bibitem [{\citenamefont {{Ono}}\ \emph {et~al.}(2018)\citenamefont {{Ono}},
  \citenamefont {{Yanase}},\ and\ \citenamefont
  {{Watanabe}}}]{Z4indicator_Tsc}%
  \BibitemOpen
  \bibfield  {author} {\bibinfo {author} {\bibfnamefont {S.}~\bibnamefont
  {{Ono}}}, \bibinfo {author} {\bibfnamefont {Y.}~\bibnamefont {{Yanase}}}, \
  and\ \bibinfo {author} {\bibfnamefont {H.}~\bibnamefont {{Watanabe}}},\
  }\href@noop {} {\bibfield  {journal} {\bibinfo  {journal} {arXiv e-prints}\
  ,\ \bibinfo {eid} {arXiv:1811.08712}} (\bibinfo {year} {2018})},\ \Eprint
  {http://arxiv.org/abs/1811.08712} {arXiv:1811.08712 [cond-mat.supr-con]}
  \BibitemShut {NoStop}%
\bibitem [{\citenamefont {Benalcazar}\ \emph {et~al.}(2017)\citenamefont
  {Benalcazar}, \citenamefont {Bernevig},\ and\ \citenamefont
  {Hughes}}]{MultipoleTI_science}%
  \BibitemOpen
  \bibfield  {author} {\bibinfo {author} {\bibfnamefont {W.~A.}\ \bibnamefont
  {Benalcazar}}, \bibinfo {author} {\bibfnamefont {B.~A.}\ \bibnamefont
  {Bernevig}}, \ and\ \bibinfo {author} {\bibfnamefont {T.~L.}\ \bibnamefont
  {Hughes}},\ }\href {\doibase 10.1126/science.aah6442} {\bibfield  {journal}
  {\bibinfo  {journal} {Science}\ }\textbf {\bibinfo {volume} {357}},\ \bibinfo
  {pages} {61} (\bibinfo {year} {2017})},\ \Eprint
  {http://arxiv.org/abs/http://science.sciencemag.org/content/357/6346/61.full.pdf}
  {http://science.sciencemag.org/content/357/6346/61.full.pdf} \BibitemShut
  {NoStop}%
\bibitem [{\citenamefont {Wang}\ \emph
  {et~al.}(2018{\natexlab{a}})\citenamefont {Wang}, \citenamefont {Liu},
  \citenamefont {Lu},\ and\ \citenamefont {Zhang}}]{ZhangMCS}%
  \BibitemOpen
  \bibfield  {author} {\bibinfo {author} {\bibfnamefont {Q.}~\bibnamefont
  {Wang}}, \bibinfo {author} {\bibfnamefont {C.-C.}\ \bibnamefont {Liu}},
  \bibinfo {author} {\bibfnamefont {Y.-M.}\ \bibnamefont {Lu}}, \ and\ \bibinfo
  {author} {\bibfnamefont {F.}~\bibnamefont {Zhang}},\ }\href {\doibase
  10.1103/PhysRevLett.121.186801} {\bibfield  {journal} {\bibinfo  {journal}
  {Phys. Rev. Lett.}\ }\textbf {\bibinfo {volume} {121}},\ \bibinfo {pages}
  {186801} (\bibinfo {year} {2018}{\natexlab{a}})}\BibitemShut {NoStop}%
\bibitem [{\citenamefont {Yan}\ \emph {et~al.}(2018)\citenamefont {Yan},
  \citenamefont {Song},\ and\ \citenamefont {Wang}}]{WangMCS}%
  \BibitemOpen
  \bibfield  {author} {\bibinfo {author} {\bibfnamefont {Z.}~\bibnamefont
  {Yan}}, \bibinfo {author} {\bibfnamefont {F.}~\bibnamefont {Song}}, \ and\
  \bibinfo {author} {\bibfnamefont {Z.}~\bibnamefont {Wang}},\ }\href {\doibase
  10.1103/PhysRevLett.121.096803} {\bibfield  {journal} {\bibinfo  {journal}
  {Phys. Rev. Lett.}\ }\textbf {\bibinfo {volume} {121}},\ \bibinfo {pages}
  {096803} (\bibinfo {year} {2018})}\BibitemShut {NoStop}%
\bibitem [{\citenamefont {Wang}\ \emph
  {et~al.}(2018{\natexlab{b}})\citenamefont {Wang}, \citenamefont {Lin},\ and\
  \citenamefont {Hughes}}]{weakHOTsc_Hughes}%
  \BibitemOpen
  \bibfield  {author} {\bibinfo {author} {\bibfnamefont {Y.}~\bibnamefont
  {Wang}}, \bibinfo {author} {\bibfnamefont {M.}~\bibnamefont {Lin}}, \ and\
  \bibinfo {author} {\bibfnamefont {T.~L.}\ \bibnamefont {Hughes}},\ }\href
  {\doibase 10.1103/PhysRevB.98.165144} {\bibfield  {journal} {\bibinfo
  {journal} {Phys. Rev. B}\ }\textbf {\bibinfo {volume} {98}},\ \bibinfo
  {pages} {165144} (\bibinfo {year} {2018}{\natexlab{b}})}\BibitemShut
  {NoStop}%
\bibitem [{\citenamefont {Schindler}\ \emph {et~al.}(2018)\citenamefont
  {Schindler}, \citenamefont {Cook}, \citenamefont {Vergniory}, \citenamefont
  {Wang}, \citenamefont {Parkin}, \citenamefont {Bernevig},\ and\ \citenamefont
  {Neupert}}]{HOTI_Neupert}%
  \BibitemOpen
  \bibfield  {author} {\bibinfo {author} {\bibfnamefont {F.}~\bibnamefont
  {Schindler}}, \bibinfo {author} {\bibfnamefont {A.~M.}\ \bibnamefont {Cook}},
  \bibinfo {author} {\bibfnamefont {M.~G.}\ \bibnamefont {Vergniory}}, \bibinfo
  {author} {\bibfnamefont {Z.}~\bibnamefont {Wang}}, \bibinfo {author}
  {\bibfnamefont {S.~S.~P.}\ \bibnamefont {Parkin}}, \bibinfo {author}
  {\bibfnamefont {B.~A.}\ \bibnamefont {Bernevig}}, \ and\ \bibinfo {author}
  {\bibfnamefont {T.}~\bibnamefont {Neupert}},\ }\href {\doibase
  10.1126/sciadv.aat0346} {\bibfield  {journal} {\bibinfo  {journal} {Science
  Advances}\ }\textbf {\bibinfo {volume} {4}} (\bibinfo {year} {2018}),\
  10.1126/sciadv.aat0346},\ \Eprint
  {http://arxiv.org/abs/http://advances.sciencemag.org/content/4/6/eaat0346.full.pdf}
  {http://advances.sciencemag.org/content/4/6/eaat0346.full.pdf} \BibitemShut
  {NoStop}%
\bibitem [{\citenamefont {Langbehn}\ \emph {et~al.}(2017)\citenamefont
  {Langbehn}, \citenamefont {Peng}, \citenamefont {Trifunovic}, \citenamefont
  {von Oppen},\ and\ \citenamefont {Brouwer}}]{MirrorHOTI_Brouwer}%
  \BibitemOpen
  \bibfield  {author} {\bibinfo {author} {\bibfnamefont {J.}~\bibnamefont
  {Langbehn}}, \bibinfo {author} {\bibfnamefont {Y.}~\bibnamefont {Peng}},
  \bibinfo {author} {\bibfnamefont {L.}~\bibnamefont {Trifunovic}}, \bibinfo
  {author} {\bibfnamefont {F.}~\bibnamefont {von Oppen}}, \ and\ \bibinfo
  {author} {\bibfnamefont {P.~W.}\ \bibnamefont {Brouwer}},\ }\href {\doibase
  10.1103/PhysRevLett.119.246401} {\bibfield  {journal} {\bibinfo  {journal}
  {Phys. Rev. Lett.}\ }\textbf {\bibinfo {volume} {119}},\ \bibinfo {pages}
  {246401} (\bibinfo {year} {2017})}\BibitemShut {NoStop}%
\bibitem [{\citenamefont {Shapourian}\ \emph {et~al.}(2018)\citenamefont
  {Shapourian}, \citenamefont {Wang},\ and\ \citenamefont {Ryu}}]{TCsc_Ryu}%
  \BibitemOpen
  \bibfield  {author} {\bibinfo {author} {\bibfnamefont {H.}~\bibnamefont
  {Shapourian}}, \bibinfo {author} {\bibfnamefont {Y.}~\bibnamefont {Wang}}, \
  and\ \bibinfo {author} {\bibfnamefont {S.}~\bibnamefont {Ryu}},\ }\href
  {\doibase 10.1103/PhysRevB.97.094508} {\bibfield  {journal} {\bibinfo
  {journal} {Phys. Rev. B}\ }\textbf {\bibinfo {volume} {97}},\ \bibinfo
  {pages} {094508} (\bibinfo {year} {2018})}\BibitemShut {NoStop}%
\bibitem [{\citenamefont {{Bultinck}}\ \emph {et~al.}(2018)\citenamefont
  {{Bultinck}}, \citenamefont {{Bernevig}},\ and\ \citenamefont
  {{Zaletel}}}]{HybridHOTsc_Zaletel}%
  \BibitemOpen
  \bibfield  {author} {\bibinfo {author} {\bibfnamefont {N.}~\bibnamefont
  {{Bultinck}}}, \bibinfo {author} {\bibfnamefont {B.~A.}\ \bibnamefont
  {{Bernevig}}}, \ and\ \bibinfo {author} {\bibfnamefont {M.~P.}\ \bibnamefont
  {{Zaletel}}},\ }\href@noop {} {\bibfield  {journal} {\bibinfo  {journal}
  {arXiv e-prints}\ ,\ \bibinfo {eid} {arXiv:1810.12963}} (\bibinfo {year}
  {2018})},\ \Eprint {http://arxiv.org/abs/1810.12963} {arXiv:1810.12963
  [cond-mat.supr-con]} \BibitemShut {NoStop}%
\bibitem [{\citenamefont {{Xu}}\ \emph {et~al.}(2019)\citenamefont {{Xu}},
  \citenamefont {{Song}}, \citenamefont {{Wang}}, \citenamefont {{Weng}},\ and\
  \citenamefont {{Dai}}}]{HOTIaxion}%
  \BibitemOpen
  \bibfield  {author} {\bibinfo {author} {\bibfnamefont {Y.}~\bibnamefont
  {{Xu}}}, \bibinfo {author} {\bibfnamefont {Z.}~\bibnamefont {{Song}}},
  \bibinfo {author} {\bibfnamefont {Z.}~\bibnamefont {{Wang}}}, \bibinfo
  {author} {\bibfnamefont {H.}~\bibnamefont {{Weng}}}, \ and\ \bibinfo {author}
  {\bibfnamefont {X.}~\bibnamefont {{Dai}}},\ }\href@noop {} {\bibfield
  {journal} {\bibinfo  {journal} {arXiv e-prints}\ ,\ \bibinfo {eid}
  {arXiv:1903.09856}} (\bibinfo {year} {2019})},\ \Eprint
  {http://arxiv.org/abs/1903.09856} {arXiv:1903.09856 [cond-mat.mtrl-sci]}
  \BibitemShut {NoStop}%
\bibitem [{\citenamefont {Zhang}\ \emph {et~al.}(2019)\citenamefont {Zhang},
  \citenamefont {Cole},\ and\ \citenamefont {Das~Sarma}}]{HingeIronsc_PRL}%
  \BibitemOpen
  \bibfield  {author} {\bibinfo {author} {\bibfnamefont {R.-X.}\ \bibnamefont
  {Zhang}}, \bibinfo {author} {\bibfnamefont {W.~S.}\ \bibnamefont {Cole}}, \
  and\ \bibinfo {author} {\bibfnamefont {S.}~\bibnamefont {Das~Sarma}},\ }\href
  {\doibase 10.1103/PhysRevLett.122.187001} {\bibfield  {journal} {\bibinfo
  {journal} {Phys. Rev. Lett.}\ }\textbf {\bibinfo {volume} {122}},\ \bibinfo
  {pages} {187001} (\bibinfo {year} {2019})}\BibitemShut {NoStop}%
\bibitem [{\citenamefont {{Zhang}}\ \emph {et~al.}(2019)\citenamefont
  {{Zhang}}, \citenamefont {{Cole}}, \citenamefont {{Wu}},\ and\ \citenamefont
  {{Das Sarma}}}]{HOTsc_hetero}%
  \BibitemOpen
  \bibfield  {author} {\bibinfo {author} {\bibfnamefont {R.-X.}\ \bibnamefont
  {{Zhang}}}, \bibinfo {author} {\bibfnamefont {W.~S.}\ \bibnamefont {{Cole}}},
  \bibinfo {author} {\bibfnamefont {X.}~\bibnamefont {{Wu}}}, \ and\ \bibinfo
  {author} {\bibfnamefont {S.}~\bibnamefont {{Das Sarma}}},\ }\href@noop {}
  {\bibfield  {journal} {\bibinfo  {journal} {arXiv e-prints}\ ,\ \bibinfo
  {eid} {arXiv:1905.10647}} (\bibinfo {year} {2019})},\ \Eprint
  {http://arxiv.org/abs/1905.10647} {arXiv:1905.10647 [cond-mat.supr-con]}
  \BibitemShut {NoStop}%
\bibitem [{\citenamefont {Zheng}\ \emph {et~al.}(2016)\citenamefont {Zheng},
  \citenamefont {Cai}, \citenamefont {Ge}, \citenamefont {Zhang}, \citenamefont
  {Liu}, \citenamefont {Lu}, \citenamefont {Zhang}, \citenamefont {Qiu},
  \citenamefont {Taniguchi}, \citenamefont {Watanabe}, \citenamefont {Jia},
  \citenamefont {Qi}, \citenamefont {Chen}, \citenamefont {Sun},\ and\
  \citenamefont {Feng}}]{DFTHybridFunc}%
  \BibitemOpen
  \bibfield  {author} {\bibinfo {author} {\bibfnamefont {F.}~\bibnamefont
  {Zheng}}, \bibinfo {author} {\bibfnamefont {C.}~\bibnamefont {Cai}}, \bibinfo
  {author} {\bibfnamefont {S.}~\bibnamefont {Ge}}, \bibinfo {author}
  {\bibfnamefont {X.}~\bibnamefont {Zhang}}, \bibinfo {author} {\bibfnamefont
  {X.}~\bibnamefont {Liu}}, \bibinfo {author} {\bibfnamefont {H.}~\bibnamefont
  {Lu}}, \bibinfo {author} {\bibfnamefont {Y.}~\bibnamefont {Zhang}}, \bibinfo
  {author} {\bibfnamefont {J.}~\bibnamefont {Qiu}}, \bibinfo {author}
  {\bibfnamefont {T.}~\bibnamefont {Taniguchi}}, \bibinfo {author}
  {\bibfnamefont {K.}~\bibnamefont {Watanabe}}, \bibinfo {author}
  {\bibfnamefont {S.}~\bibnamefont {Jia}}, \bibinfo {author} {\bibfnamefont
  {J.}~\bibnamefont {Qi}}, \bibinfo {author} {\bibfnamefont {J.-H.}\
  \bibnamefont {Chen}}, \bibinfo {author} {\bibfnamefont {D.}~\bibnamefont
  {Sun}}, \ and\ \bibinfo {author} {\bibfnamefont {J.}~\bibnamefont {Feng}},\
  }\href {\doibase 10.1002/adma.201600100} {\bibfield  {journal} {\bibinfo
  {journal} {Adv. Mater.}\ }\textbf {\bibinfo {volume} {28}},\ \bibinfo {pages}
  {4845} (\bibinfo {year} {2016})}\BibitemShut {NoStop}%
\bibitem [{\citenamefont {Klemm}\ \emph {et~al.}(1975)\citenamefont {Klemm},
  \citenamefont {Luther},\ and\ \citenamefont {Beasley}}]{Hcs_SOCrate}%
  \BibitemOpen
  \bibfield  {author} {\bibinfo {author} {\bibfnamefont {R.~A.}\ \bibnamefont
  {Klemm}}, \bibinfo {author} {\bibfnamefont {A.}~\bibnamefont {Luther}}, \
  and\ \bibinfo {author} {\bibfnamefont {M.~R.}\ \bibnamefont {Beasley}},\
  }\href {\doibase 10.1103/PhysRevB.12.877} {\bibfield  {journal} {\bibinfo
  {journal} {Phys. Rev. B}\ }\textbf {\bibinfo {volume} {12}},\ \bibinfo
  {pages} {877} (\bibinfo {year} {1975})}\BibitemShut {NoStop}%
\bibitem [{Note1()}]{Note1}%
  \BibitemOpen
  \bibinfo {note} {$M_x$ can be transformed back to the conventional mirror
  symmetry by shifting the mirror plane away from inversion
  center.}\BibitemShut {Stop}%
\bibitem [{\citenamefont {Muechler}\ \emph {et~al.}(2016)\citenamefont
  {Muechler}, \citenamefont {Alexandradinata}, \citenamefont {Neupert},\ and\
  \citenamefont {Car}}]{muechler_2016prx}%
  \BibitemOpen
  \bibfield  {author} {\bibinfo {author} {\bibfnamefont {L.}~\bibnamefont
  {Muechler}}, \bibinfo {author} {\bibfnamefont {A.}~\bibnamefont
  {Alexandradinata}}, \bibinfo {author} {\bibfnamefont {T.}~\bibnamefont
  {Neupert}}, \ and\ \bibinfo {author} {\bibfnamefont {R.}~\bibnamefont
  {Car}},\ }\href {\doibase 10.1103/PhysRevX.6.041069} {\bibfield  {journal}
  {\bibinfo  {journal} {Phys. Rev. X}\ }\textbf {\bibinfo {volume} {6}},\
  \bibinfo {pages} {041069} (\bibinfo {year} {2016})}\BibitemShut {NoStop}%
\bibitem [{\citenamefont {{Ok}}\ \emph {et~al.}(2018)\citenamefont {{Ok}},
  \citenamefont {{Muechler}}, \citenamefont {{Di Sante}}, \citenamefont
  {{Sangiovanni}}, \citenamefont {{Thomale}},\ and\ \citenamefont
  {{Neupert}}}]{WTe2_tbSOC}%
  \BibitemOpen
  \bibfield  {author} {\bibinfo {author} {\bibfnamefont {S.}~\bibnamefont
  {{Ok}}}, \bibinfo {author} {\bibfnamefont {L.}~\bibnamefont {{Muechler}}},
  \bibinfo {author} {\bibfnamefont {D.}~\bibnamefont {{Di Sante}}}, \bibinfo
  {author} {\bibfnamefont {G.}~\bibnamefont {{Sangiovanni}}}, \bibinfo {author}
  {\bibfnamefont {R.}~\bibnamefont {{Thomale}}}, \ and\ \bibinfo {author}
  {\bibfnamefont {T.}~\bibnamefont {{Neupert}}},\ }\href@noop {} {\bibfield
  {journal} {\bibinfo  {journal} {arXiv e-prints}\ ,\ \bibinfo {eid}
  {arXiv:1811.00551}} (\bibinfo {year} {2018})},\ \Eprint
  {http://arxiv.org/abs/1811.00551} {arXiv:1811.00551 [cond-mat.mes-hall]}
  \BibitemShut {NoStop}%
\bibitem [{Note2()}]{Note2}%
  \BibitemOpen
  \bibinfo {note} {Ref. \protect \rev@citealpnum {tbmodelmoreSOC} showed that
  more spin-orbit coupling terms are required for a better fit to the
  experimentally data. We nonetheless expect our results to change only
  quantitatively.}\BibitemShut {Stop}%
\bibitem [{\citenamefont {Sigrist}(2005)}]{ScLectureSigrist}%
  \BibitemOpen
  \bibfield  {author} {\bibinfo {author} {\bibfnamefont {M.}~\bibnamefont
  {Sigrist}},\ }\href {\doibase 10.1063/1.2080350} {\bibfield  {journal}
  {\bibinfo  {journal} {AIP Conference Proceedings}\ }\textbf {\bibinfo
  {volume} {789}},\ \bibinfo {pages} {165} (\bibinfo {year} {2005})},\ \Eprint
  {http://arxiv.org/abs/https://aip.scitation.org/doi/pdf/10.1063/1.2080350}
  {https://aip.scitation.org/doi/pdf/10.1063/1.2080350} \BibitemShut {NoStop}%
\bibitem [{Note3()}]{Note3}%
  \BibitemOpen
  \bibinfo {note} {For numerical convenience we take the self-consistent $B_u$
  symmetry solution and multiply by 10, so that the resulting superconducting
  gaps are always much larger than any finite-size gaps of the normal bulk or
  edge states for tractable lattice sizes.}\BibitemShut {Stop}%
\bibitem [{Note4()}]{Note4}%
  \BibitemOpen
  \bibinfo {note} {The superscript denotes a different spin orientation from
  that of $B_u^{(')}$.}\BibitemShut {Stop}%
\bibitem [{Note5()}]{Note5}%
  \BibitemOpen
  \bibinfo {note} {There could be more recipes to achieve such a
  phase.}\BibitemShut {Stop}%
\bibitem [{\citenamefont {Fu}\ and\ \citenamefont {Kane}(2007)}]{FuKaneZ2}%
  \BibitemOpen
  \bibfield  {author} {\bibinfo {author} {\bibfnamefont {L.}~\bibnamefont
  {Fu}}\ and\ \bibinfo {author} {\bibfnamefont {C.~L.}\ \bibnamefont {Kane}},\
  }\href {\doibase 10.1103/PhysRevB.76.045302} {\bibfield  {journal} {\bibinfo
  {journal} {Phys. Rev. B}\ }\textbf {\bibinfo {volume} {76}},\ \bibinfo
  {pages} {045302} (\bibinfo {year} {2007})}\BibitemShut {NoStop}%
\bibitem [{\citenamefont {Kane}\ and\ \citenamefont
  {Mele}(2005{\natexlab{b}})}]{KaneMele_Z2}%
  \BibitemOpen
  \bibfield  {author} {\bibinfo {author} {\bibfnamefont {C.~L.}\ \bibnamefont
  {Kane}}\ and\ \bibinfo {author} {\bibfnamefont {E.~J.}\ \bibnamefont
  {Mele}},\ }\href {\doibase 10.1103/PhysRevLett.95.146802} {\bibfield
  {journal} {\bibinfo  {journal} {Phys. Rev. Lett.}\ }\textbf {\bibinfo
  {volume} {95}},\ \bibinfo {pages} {146802} (\bibinfo {year}
  {2005}{\natexlab{b}})}\BibitemShut {NoStop}%
\bibitem [{\citenamefont {Geier}\ \emph {et~al.}(2018)\citenamefont {Geier},
  \citenamefont {Trifunovic}, \citenamefont {Hoskam},\ and\ \citenamefont
  {Brouwer}}]{ExtrinsicHOTsc}%
  \BibitemOpen
  \bibfield  {author} {\bibinfo {author} {\bibfnamefont {M.}~\bibnamefont
  {Geier}}, \bibinfo {author} {\bibfnamefont {L.}~\bibnamefont {Trifunovic}},
  \bibinfo {author} {\bibfnamefont {M.}~\bibnamefont {Hoskam}}, \ and\ \bibinfo
  {author} {\bibfnamefont {P.~W.}\ \bibnamefont {Brouwer}},\ }\href {\doibase
  10.1103/PhysRevB.97.205135} {\bibfield  {journal} {\bibinfo  {journal} {Phys.
  Rev. B}\ }\textbf {\bibinfo {volume} {97}},\ \bibinfo {pages} {205135}
  (\bibinfo {year} {2018})}\BibitemShut {NoStop}%
\bibitem [{\citenamefont {{Skurativska}}\ \emph {et~al.}(2019)\citenamefont
  {{Skurativska}}, \citenamefont {{Neupert}},\ and\ \citenamefont
  {{Fischer}}}]{Indicator_Fischer}%
  \BibitemOpen
  \bibfield  {author} {\bibinfo {author} {\bibfnamefont {A.}~\bibnamefont
  {{Skurativska}}}, \bibinfo {author} {\bibfnamefont {T.}~\bibnamefont
  {{Neupert}}}, \ and\ \bibinfo {author} {\bibfnamefont {M.~H.}\ \bibnamefont
  {{Fischer}}},\ }\href@noop {} {\bibfield  {journal} {\bibinfo  {journal}
  {arXiv e-prints}\ ,\ \bibinfo {eid} {arXiv:1906.11267}} (\bibinfo {year}
  {2019})},\ \Eprint {http://arxiv.org/abs/1906.11267} {arXiv:1906.11267
  [cond-mat.supr-con]} \BibitemShut {NoStop}%
\bibitem [{\citenamefont {{Lau}}\ \emph {et~al.}(2018)\citenamefont {{Lau}},
  \citenamefont {{Ray}}, \citenamefont {{Varjas}},\ and\ \citenamefont
  {{Akhmerov}}}]{tbmodelmoreSOC}%
  \BibitemOpen
  \bibfield  {author} {\bibinfo {author} {\bibfnamefont {A.}~\bibnamefont
  {{Lau}}}, \bibinfo {author} {\bibfnamefont {R.}~\bibnamefont {{Ray}}},
  \bibinfo {author} {\bibfnamefont {D.}~\bibnamefont {{Varjas}}}, \ and\
  \bibinfo {author} {\bibfnamefont {A.}~\bibnamefont {{Akhmerov}}},\
  }\href@noop {} {\bibfield  {journal} {\bibinfo  {journal} {arXiv e-prints}\
  ,\ \bibinfo {eid} {arXiv:1812.05693}} (\bibinfo {year} {2018})},\ \Eprint
  {http://arxiv.org/abs/1812.05693} {arXiv:1812.05693 [cond-mat.mes-hall]}
  \BibitemShut {NoStop}%
\end{thebibliography}
\end{document}